\documentclass[prd,twocolumn,amsmath,amssymb,floatfix,superscriptaddress,nofootinbib]{revtex4-1}

\usepackage{graphicx}
\usepackage{scrextend}
\usepackage{amssymb}
\usepackage{amsmath}
\usepackage{bm}
\usepackage{color}
\usepackage{multirow}
\usepackage{cprotect}

\DeclareMathAlphabet\mathbfcal{OMS}{cmsy}{b}{n}
\definecolor{darkgreen}{RGB}{50,150,0}
\definecolor{purple}{cmyk}{0.5,1.0,0,0}

\def\edth{\;\raise1.0pt\hbox{$'$}\hskip-6pt\partial}
\def\baredth{\;\overline{\raise1.0pt\hbox{$'$}\hskip-6pt
\partial}}

\newcommand{\SSC}{{\rm SSC}}
\newcommand{\ILC}{{\rm ILC}}
\def\be{\begin{equation}}
\def\ee{\end{equation}}
\def\ben{\begin{equation} \nonumber}
\def\een{\end{equation}}
\def\ban{\begin{eqnarray*}}
\def\ean{\end{eqnarray*}}
\def\ba{\begin{eqnarray}}
\def\ea{\end{eqnarray}}
\def\({\left(}
\def\){\right)}

\newcommand{\Cov}{\mathrm{Cov}}
\newcommand{\PP}{{\phi\phi}}

\definecolor{ultramarine}{rgb}{0.07, 0.04, 0.56}
\definecolor{cadmiumgreen}{rgb}{0.0, 0.42, 0.24}
\definecolor{indigo(dye)}{rgb}{0.0, 0.25, 0.42}
\usepackage[linktocpage=true]{hyperref}
\hypersetup{
colorlinks=true,
citecolor=ultramarine,
linkcolor=cadmiumgreen,
urlcolor=indigo(dye),
pdfauthor={},
pdftitle={},
pdfsubject={}
}

\begin{document}

\title{Lensing covariance on cut sky and SPT-Planck lensing tensions}

\author{Pavel Motloch}
\affiliation{Canadian Institute for Theoretical Astrophysics, 60 St. George St.,
Toronto, Ontario M5S 3H8, Canada}
\affiliation{Kavli Institute for Cosmological Physics, Department of Physics, University of Chicago, Chicago, Illinois 60637, U.S.A}

\author{Wayne Hu}
\affiliation{Kavli Institute for Cosmological Physics, Department of Astronomy \& Astrophysics,  Enrico Fermi Institute, University of Chicago, Chicago, Illinois 60637, U.S.A}
\begin{abstract}
\noindent
We investigate correlations induced by  gravitational lensing  on simulated cosmic microwave background data of experiments with an incomplete sky
coverage and their effect on inferences from the South Pole Telescope data. These correlations agree well with the theoretical expectations, given by the sum of
super-sample and intra-sample lensing terms, with only a typically negligible $\sim$ 5\% discrepancy in the amplitude
of the super-sample lensing effect. Including these effects we find that lensing
constraints are in $3.0\sigma$ or
 $2.1\sigma$ tension between the SPT polarization measurements and Planck temperature
or lensing reconstruction constraints respectively. If the lensing-induced covariance effects are
neglected, the significance of these tensions increases to $3.5\sigma$ or $2.5\sigma$.
{Using the standard scaling parameter $A_L$ 
substantially
underestimates the significance of the tension once other parameters are marginalized over.} 
By parameterizing the
super-sample lensing through the mean convergence in the SPT footprint, we 
find a hint of underdensity in the SPT region.  We also constrain extra sharpening of the CMB acoustic peaks due to missing smoothing of the
peaks by super-sample lenses at a level that is much smaller than the lens sample variance. Finally, we
extend  the usual ``shift in the means'' statistic for evaluating tensions to
non-Gaussian posteriors, {generalize an approach to extract correlation  modes from noisy simulated
covariance matrices, and present a treatment of 
correlation modes not as data covariances but as auxiliary model parameters.}
\end{abstract}

\maketitle

\section{Introduction}
\label{sec:intro}

Cosmic microwave background (CMB) measurements \cite{Spergel:2003cb, Adam:2015rua} have
been instrumental in confirming the $\Lambda$ cold dark matter model ($\Lambda$CDM) as the
standard model of cosmology and in constraining its parameters. Gravitational lensing of the
CMB (see \cite{Lewis:2006fu} for a review), recently measured with high
significance by a number of experiments \cite{Smith:2007rg, Hirata08,Hanson:2013hsb,
Das:2011ak, Keisler:2011aw, Planck2013XVII, Keisler:2015hfa,Ade:2015zua, Array:2016afx,
Sherwin:2016tyf, Aghanim:2018oex}, is a secondary effect that allows us to break geometric degeneracy in
the CMB data and constrain the low redshift Universe parameters, such as properties of the
dark energy and the sum of neutrino masses. Upcoming CMB experiments \cite{Benson:2014qhw,
Henderson:2015nzj, Wu:2016hul, Ade:2018sbj} promise to greatly improve on these
measurements and make unprecedented measurements of the low redshift physics in the linear regime.

With the increasing precision of the measurements, it is necessary to dedicate increasing
scrutiny to  
{subtle  effects that have been omitted so far in most analyses. }
One such effect is the
non-Gaussian correlations induced in the CMB data by the gravitational lensing
\cite{BenoitLevy:2012va, Manzotti:2014wca}, reflecting the stochastic nature of the
gravitational lensing potential $\phi$. These correlations have been investigated on
simulations for an idealized full-sky experiment \cite{BenoitLevy:2012va,
Peloton:2016kbw}. However, in reality all CMB experiments {can only utilize the information on} a portion of the sky. It
is thus timely to investigate lensing-induced covariances for a cut sky experiments and
check {their theoretical description} on simulations. Such a study is presented in the first part
of this work.

Then we focus on the CMB polarization measurements from the South Pole Telescope (SPT) to
better understand how the lensing-induced covariance terms manifests on the
cosmological parameter level and how they affect information extracted from the
lensing potential.  For the latter, these effects are already important for SPT data.

Lastly, while the standard cosmological model is a very good description of the
experimental data, there are several tensions that can potentially signal presence of new
physics \cite{Raveri:2018wln, Riess:2018aaa, Kohlinger:2017sxk}. One of the problems is an
anomalously high amount of lensing detected through the smoothing of the acoustic peaks in
the Planck temperature power spectra \cite{Ade:2015xua, Aghanim:2016sns,
Addison:2015wyg}. Using a novel technique that allows a direct comparison of gravitational
lensing constraints obtained from various data sets
\cite{Motloch:2016zsl,Motloch:2017rlk}, it is possible to check to what extent the SPT
lensing measurements agree with the lensing constraints from Planck, as we do in the
final part of this work.

The outline of the paper is as follows. In \S\,\ref{sec:lic_investigations} we present our
numerical simulations of lensed CMB data and study their covariances as determined by cut
sky experiments. We provide several technical details as Appendices, quantifying agreement
between the simulated covariance matrices and theoretical expectations of the
lensing-induced covariance terms in Appendix~\ref{sec:app_agreement} and constraining
extra sharpening of the CMB acoustic peaks due to missing smoothing by the super-sample
lenses in Appendix~\ref{sec:app_sharpening}. In \S\,\ref{sec:analysis_details} we present
the data sets used in this paper and discuss the details of their analysis. In
\S\,\ref{sec:lic_in_spt_teee} we conduct a case study of the effect of the lensing-induced
covariance terms in the SPTpol polarization likelihood, especially in terms of how they
affect cosmological parameter constraints. Finally, in \S\,\ref{sec:tensions} we compare
constraints on gravitational lensing from various SPT and Planck data sets using a
generalization of the standard ``shift in the means'' statistic, which we present in
Appendix~\ref{sec:app_significance}. We conclude in \S\,\ref{sec:discussion}.

\section{Lensing induced covariance in cut sky simulations}
\label{sec:lic_investigations}

In this section we describe our simulations of CMB experiments with an incomplete sky
coverage and briefly summarize the standard pseudo-$C_\ell$ method, before we present our
results on CMB power spectra covariances and compare them against theoretical expectations.

\subsection{Simulations}
\label{sec:simulations}

To simulate lensed CMB data we use the publicly available code
Lenspix\footnote{https://github.com/cmbant/lenspix} \cite{Lewis:2005tp} with unlensed CMB
power spectra calculated by CAMB\footnote{http://camb.info} \cite{Lewis:1999bs}.

The fiducial cosmological model used to calculate the simulated CMB data is the 
best fit flat $\Lambda$CDM cosmological model, determined from the 2015 Planck temperature
and low-$\ell$ polarization likelihoods assuming no primordial tensor modes and minimal
mass neutrinos ($\sum m_\nu = 60\, \mathrm{meV}$).  To reflect the updated results on the
optical depth to recombination $\tau$ from \cite{Adam:2016hgk}, we set $\tau$ to the value
from that work and decrease $A_s$ to keep $A_s e^{-2\tau}$ constant. 

The six parameters of the $\Lambda$CDM model are:  $\Omega_b h^2$, the physical baryon
density; $\Omega_c h^2$, the physical cold dark matter density; $n_s$, the tilt of the
scalar power spectrum; $\ln A_s$, its log amplitude at $k=0.05$ Mpc$^{-1}$; $\tau$ the
optical depth through reionization, and $\theta_*$, the angular scale of the sound horizon
at recombination. Their fiducial values considered in this work are listed in
Table~\ref{tab:fiducial}. 

\begin{table}
\caption{Fiducial $\Lambda$CDM parameters used in this work.\footnote{In $\Lambda$CDM, these parameters also imply a Hubble constant
of $h=0.6733$.}
}
\label{tab:fiducial}
\begin{tabular}{c|c}
\hline\hline
Parameter & Fiducial value\\
\hline
100 $\theta_*$ & 1.041 \\
$\Omega_c h^2$ & $0.1197$ \\
$\Omega_b h^2$ & $0.02223$ \\
$n_s$ & $0.9658$\\
$\ln(10^{10} A_s)$ & $3.049$  \\
$\tau$ &  $0.058$\\
\hline\hline
\end{tabular}
\end{table}

\subsection{Pseudo-$C_\ell$}
\label{sec:pseudocl}

In this section we briefly summarize the standard pseudo-$C_\ell$ approach of analyzing
the cut-sky CMB data \cite{Hivon:2001jp,Brown:2004jn}.

The part of the sky observed by a finite survey can be described by a window function
(also called mask) $w(\hat n)$,
that is zero outside of the observed region. Inside, $w$ can be chosen to attain values between
zero and one, for example to reduce ringing in the Fourier space.

Effectively, such experiments measure the fluctuations of the underlying CMB temperature
$T$ and Stokes $Q$ and $U$ parameters windowed,
\ba
	T_w(\hat n) &=& w(\hat n) T(\hat n),\nonumber\\
	Q_w(\hat n) &=& w(\hat n) Q(\hat n),\nonumber\\
	U_w(\hat n) &=& w(\hat n) U(\hat n) .
\ea
As usual, it is possible to transfer from the masked fields $(T_w, Q_w, U_w)$ to the spin
and parity eigenstates $(T_w, E_w, B_w)$. Their power spectra 
\be
	\hat C_{w,\ell}^{XY} = \sum_m \frac{ X_{w,\ell m}^* Y_{w,\ell m}}{2\ell+1}
\ee
are called pseudo-$C_\ell$ power spectra.
Here we use $XY, WZ$ to denote elements from $\{TT, TE, EE,
BB\}$.

Given a statistically isotropic underlying CMB sky,
\be
\langle X_{\ell' m'}^* Y_{\ell m} \rangle = \delta_{\ell \ell'} \delta_{m m'} C_\ell^{XY},
\ee
the ensemble average of the pseudo-$C_\ell$ power spectra are linearly related to the
power spectra of the underlying CMB $C_{\ell}^{XY}$ as
\ba
	C_{w,\ell}^{XY} 
	= \sum_{WZ, \ell'} M_{\ell \ell'}^{XY,WZ} 
	C_{\ell'}^{WZ} 
	.
\ea
Analytical expressions for the mode coupling
matrices $M_{\ell \ell'}^{XY,WZ}$ can be found for example in \cite{Brown:2004jn}.   The mask
mixes $E$ and $B$ modes but in this work, we neglect the information in the lensed $C_\ell^{BB}$ spectra
and focus only on the $TT,
TE$ and $EE$ power spectra.
{Without  $C_\ell^{BB}$,}
$M^{XY,WZ}_{\ell\ell'}$ is  diagonal in $XY$, so from this point forward below we use a
shorthand notation $M^{XY}_{\ell\ell'}$ with $XY$ denoting $\{TT, TE, EE\}$.

For cut-sky experiments it is not possible to invert the mode coupling matrices for every $\ell$ and so 
we bin the power spectra in $\ell$ as
\be
\mathcal{C}_b^{XY} = \sum_\ell P_{b \ell} C_\ell^{XY},
\ee
using the binning operator
\ba
	P_{b\ell} &=& \frac{1}{\Delta \ell_b}\times
	\begin{cases}
	 \frac{\ell(\ell+1)}{2\pi }, & \mathrm{if\ }  \ell_b  \leq \ell <
	\ell_{b+1}\\
	 0, & \mathrm{otherwise}
	 \end{cases} 
	  ,
\ea
where the minimum multipole in the first bin is $\ell_1=2$.  Unless otherwise noted, we use fixed bin widths
 $\Delta \ell_b \equiv \ell_{b+1}
-\ell_b = 50$  in this work.  
The reciprocal operator reads
\ba
	Q_{\ell b} &=& 
	\begin{cases} \frac{2\pi}{\ell(\ell+1)}, &  \mathrm{if\ } \ell_b  \leq \ell <
	\ell_{b+1}\\
	0, & \mathrm{otherwise} 
	\end{cases}.
\ea
We do not include instrumental noise, beam or filtering in our simulations. 

Under these
conditions, the unbiased estimator of the underlying binned true power spectra
$\mathcal{C}_b$ is
\be
\label{unbiased_estimator}
	\hat{\mathcal{C}}_b^{XY} = \sum_{b'\ell}\(K^{XY}\)^{-1}_{bb'}P_{b' \ell} {\hat C}_{w,\ell}^{XY} ,
\ee
where
\be
	K^{XY}_{bb'} = \sum_{\ell\ell'} P_{b \ell}M^{XY}_{\ell \ell'}Q_{\ell' b'} .
\ee

For future convenience we also define the operator
\be
\mathcal{U}^{XY}_{b\ell} = \sum_{b' \ell'}\(K^{XY}\)^{-1}_{bb'}P_{b'\ell'} M^{XY}_{\ell'\ell} 
\ee
that enables comparison of full sky power spectra against simulation results.

In this work we investigate five different window functions. Four of them are circular
caps of sizes $150\,\mathrm{deg}^2$, $250\,\mathrm{deg}^2$, $500\,\mathrm{deg}^2$ and
$1000\,\mathrm{deg}^2$. The fifth is a $500\, \mathrm{deg}^2$ rectangular patch of sky
representing the SPTpol footprint, spanning 4 hr of right ascension, from 22~hr to 2~hr,
and $15^\circ$ of declination, from $-65^\circ$ to $-50^\circ$. All window functions have
been apodized by a $15'$ cosine taper to reduce ringing in  Fourier space. 

\subsection{Power spectra covariance}

\label{sec:mcmc_results}

\begin{figure*}
\center
\includegraphics[width = 0.49 \textwidth]{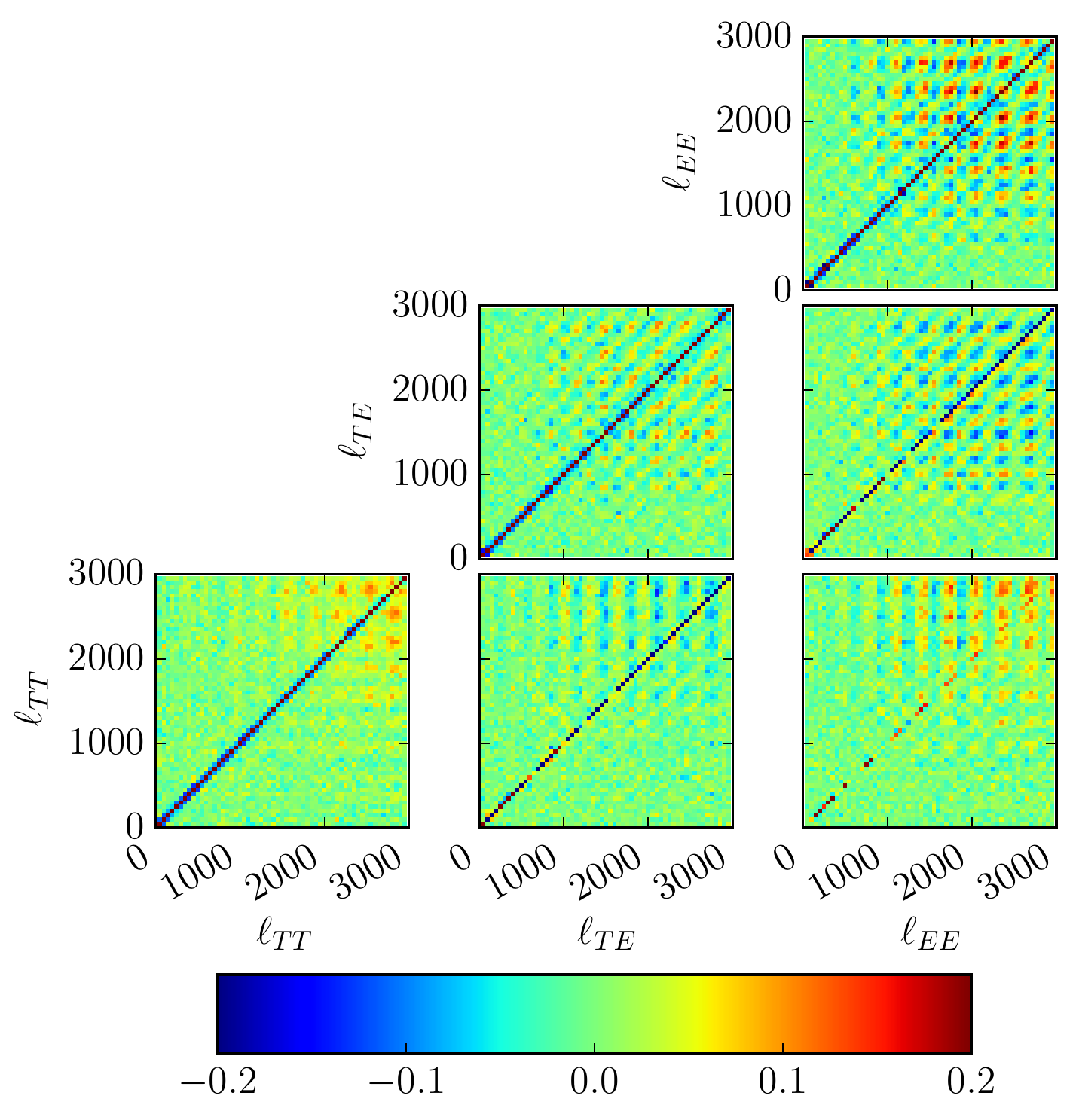}
\includegraphics[width = 0.49 \textwidth]{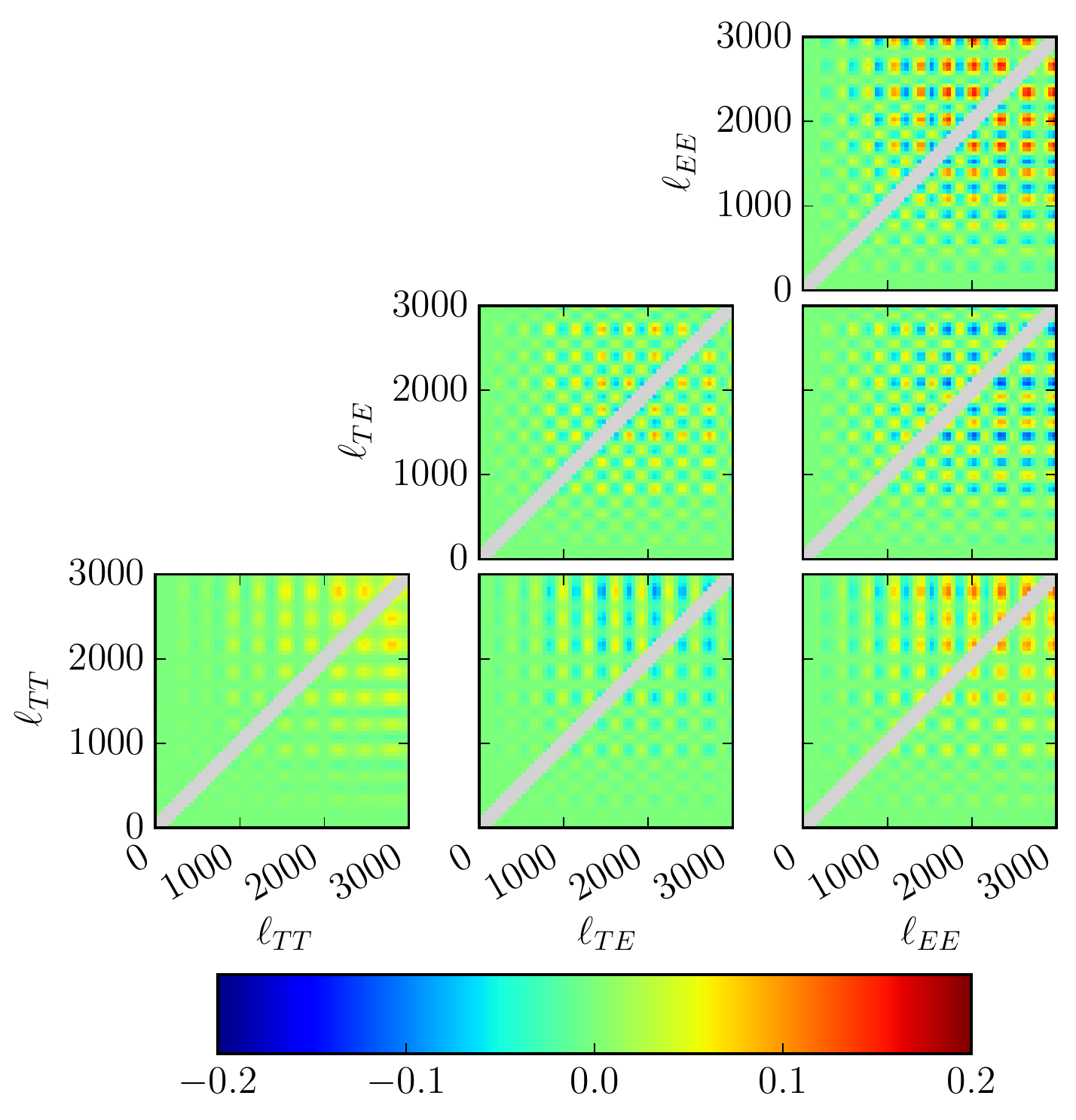}
\caption{
Left: Correlation matrix of the unbiased estimates of the binned power spectra $\hat{\mathcal{C}}_b^{XY}$
calculated from the simulations with the rectangular SPT-like window function.
Right: Theoretical expectation for the lensing-induced contributions to the correlation
matrix, {$R^{XY,WZ}_{\mathrm{(theory)bb'}}$, given by the} sum of the SSC
\eqref{ssc} and ILC \eqref{ilc} contributions. The gray region hides elements dominated by
the Gaussian terms and the window function effects.
}
\label{fig:mcmc_cov_matrix}
\end{figure*}

Using 2400 simulated CMB skies from \S\,\ref{sec:simulations} and the windows from \S\,\ref{sec:pseudocl}, we estimate the underlying binned power spectrum $ \hat{ \mathcal{C}}_b^{XY}$
using \eqref{unbiased_estimator}.   Note that for each simulated sky we extract an
estimator for
each window.   These estimators are nearly independent as we place these windows in vastly separated regions
of sky.
 For each window function, we then calculate the corresponding
covariance matrix 
\begin{equation}
\Cov_{bb'}^{XY,WZ} = \langle \hat{ \mathcal{C}}_b^{XY} \hat{ \mathcal{C}}_{b'}^{WZ} \rangle
-\langle \hat {\mathcal{C}}_b^{XY} \rangle \langle \hat {\mathcal{C}}_{b'}^{WZ} \rangle.
\end{equation}
The correlation matrix 
\be
\label{correlation}
	{\hat R^{XY,WZ}_{bb'}} =
	\frac{\Cov^{XY,WZ}_{bb'}}
	{\sqrt{\Cov_{bb}^{XY,XY}\Cov_{b'b'}^{WZ,WZ}}}
\ee
obtained for the SPTpol rectangular window is shown in Fig.~\ref{fig:mcmc_cov_matrix};
correlation matrices derived using the other windows show qualitatively similar features. 

The covariance is composed of several contributions. The diagonals in $\Cov^{TT,TE},
\Cov^{TT,EE}$ and $\Cov^{TE,EE}$ are dominated by the usual Gaussian contributions, while the
anti-correlated band around the main diagonal reflects the mode couplings due to the
window function. Both of these effects are present also for the unlensed CMB fields.

As has been previously argued \cite{Manzotti:2014wca}, gravitational lensing by lenses
larger than the survey footprint leads to coherent (de)magnification,  increasing
or decreasing the observed angular scales within the footprint. Its effect on the power spectra is thus largely
degenerate with a change in $\theta_*$, the angular scale of the acoustic peaks. This
effect and the ensuing covariance, called super-sample covariance (SSC), can be modeled as 
\be
\label{ssc}
	\Cov^{XY,WZ}_{(\SSC)bb'} = 
	\sum_{\ell\ell'}
	\mathcal{U}^{XY}_{b\ell}
	\frac{\partial \ell^2 C^{XY}_\ell}{\partial \ln \ell}
	\frac{\sigma_\kappa^2}{\ell^2 {\ell'}^2}
	\frac{\partial {\ell'}^2 C^{WZ}_{\ell'}}{\partial \ln \ell'}
	\mathcal{U}^{WZ}_{b'\ell'} .
\ee
Here $\sigma_\kappa^2$ is variance of the convergence field $\kappa = - \nabla^2
\phi/2$ in the footprint,
\be
\label{sigma_kappa_2}
	\sigma_\kappa^2 = \frac{1}{A^2}\sum_{LM} \left|w_{LM}\right|^2 \frac{L^2(L+1)^2}{4}
	C_L^{\phi\phi} ,
\ee
where
$A$ is the sky area (in radians) covered by the survey,
\be
	A = \int \mathrm{d}\hat n\, w(\hat n),
\ee
and $w_{LM}$ are the spherical harmonic
coefficients of the window function. The factors $\mathcal{U}$ are added on top of the
standard expression from \cite{Manzotti:2014wca} to represent effects of the window
function and the subsequent de-biasing. We find that in the simulated covariances, SSC is
the dominant effect induced by the gravitational lensing and corresponds to the
checkerboard pattern visible in Fig.~\ref{fig:mcmc_cov_matrix} (see also
Fig.~\ref{fig:ueli_eigenvalues} in Appendix~\ref{sec:app_alternative_method}).

Finally, fluctuations of lenses within the observed footprint also correlate CMB data
\cite{BenoitLevy:2012va} and the intra-sample lensing covariance (ILC) they induce is
given by
\ba
\label{ilc}
	\lefteqn
	{\Cov^{XY,WZ}_{(\ILC)bb'}} \\
	&= \dfrac{4\pi}{A}
	\sum\limits_{L\ell\ell'}
	\mathcal{U}^{XY}_{b\ell}
	 \left[
\,
	\dfrac{\partial C^{XY}_\ell}{\partial C_L^{\phi\phi}}
	\dfrac{2(C_L^{\phi\phi})^2}{(2L+1)}
	\dfrac{\partial C^{WZ}_{\ell'}}{\partial C_L^{\phi\phi}}
\,
	\right] 
	\mathcal{U}^{WZ}_{b'\ell'}.\nonumber
\ea
It represents the correlation caused by the common dependence of the CMB power spectra
on the stochastic lensing power. 
The inverse proportionality to the sky area $A$ reflects the fact that
due to a smaller number of measured modes, lensing power shows larger sample variance on
smaller patches.

{In  Fig.~\ref{fig:mcmc_cov_matrix} (right panel)
 we show the expected contribution of the lensing-induced terms, given by the sum of
the SSC \eqref{ssc} and ILC \eqref{ilc} contributions, to the correlation matrix  for the rectangular window
function, 
\be
\label{model_correlation}
	R^{XY,WZ}_{(\mathrm{theory})bb'} =
	\frac{\Cov^{XY,WZ}_{(\mathrm{ILC})bb'} + \Cov^{XY,WZ}_{(\mathrm{SSC})bb'}}
	{\sqrt{\Cov_{bb}^{XY,XY}\Cov_{b'b'}^{WZ,WZ}}} .
\ee
Notice we divide by the full covariance obtained from the simulations, as we do not model
the Gaussian terms explicitly and focus only on the lensing-induced terms away from the
diagonal.}

In Appendix~\ref{sec:app_agreement} we quantify the agreement between the
lensing-induced effects in the simulated covariance matrices $\Cov^{XY,WZ}_{bb'}$ and
their theoretical expectations. Using template fitting, in
Appendix~\ref{sec:app_template_fitting} we find that theoretical expectations match the
simulations reasonably well, with the amplitude of the SSC term underestimated in the model
by about $\sim$ 5 \%.  In Appendix~\ref{sec:app_alternative_method} we introduce an
alternative quantification approach, based on an idea presented in
\cite{HarnoisDeraps:2011rb}. This method gives consistent results with the template
fitting estimation.  

Finally, in cut-sky experiments we expect to see slightly sharper acoustic peaks than
in a full sky experiment. As explained above, gravitational lenses larger than the
footprint lead to a coherent shift of the angular scale. A full sky experiment contains
many lenses of such size, some of them locally magnifying while some of them locally
demagnifying the CMB fields.  Averaging over all of these lenses then leads to smoothing
of the peaks.  Because a cut sky experiment observes only one such lens, this averaging
does not happen and we in principle expect sharper peaks. In
Appendix~\ref{sec:app_sharpening} we investigate this effect on simulations and find that
for the windows that we consider it is negligible even for cosmic
variance limited CMB experiments.

\section{MCMC analysis details}
\label{sec:analysis_details}

In this section we provide details about the {Markov Chain Monte Carlo (MCMC)}
analyses we perform to find constraints of gravitational lensing from several data sets.
We start by summarizing the data  used in this paper, after which we introduce
the technique used to obtain direct measurements of the gravitational lensing potential
from the CMB data. We conclude with a few technical details about sampling of the
posterior probability distribution.

\subsection{Data}

We compare SPT lensing constraints with those from the Planck
satellite\footnote{http://pla.esac.esa.int/pla/}, derived from their 
2018 gravitational lensing reconstruction likelihood (\verb|Planck PP|)
\cite{Aghanim:2018oex} and the 2015 temperature likelihood
(\verb|Planck TT|) \cite{Aghanim:2015xee}.  
As \verb|Planck TT| is not able to measure the optical depth through
reionization $\tau$, we supplement it with a Gaussian prior on $\tau$ centered on 0.058
and with width 0.02 \cite{Adam:2016hgk}. We do not use the latest Planck parameter values
\cite{Aghanim:2018eyx} that were not available at the inception of this work; we checked
that the tensions between the data sets discussed in \S\,\ref{sec:tensions} are insensitive
to the details of the $\tau$ prior (see also \cite{Motloch:2018pjy}).

We use the publicly available SPT
likelihoods\footnote{https://lambda.gsfc.nasa.gov/product/spt/}: the SPT-SZ measurement of
$C_\ell^{TT}$ from 2500 deg${}^2$ of the sky \cite{Keisler:2011aw}, lensing
reconstruction likelihood based on the same data combined with the Planck temperature
measurement \cite{Simard:2017xtw} and the SPTpol measurements of $C_\ell^{TE}$ and 
$C_\ell^{EE}$ in a 500 deg${}^2$ patch \cite{Henning:2017nuy}. Below, we denote
these likelihoods as \verb|SPT TT|, \verb|SPT PP| and \verb|SPT TEEE|. We supplement
\verb|SPT TT| and \verb|SPT TEEE| with the same $\tau$ prior that we use for
\verb|Planck TT|.

In the sections below we show that lensing-induced covariance effects,
discussed in the previous section, are important enough to affect results derived from the
SPT polarization measurements. In \S\,\ref{sec:lic_in_spt_teee} we discuss ways how to
modify the \verb|SPT TEEE| likelihood to properly include these effects.

\subsection{Parameterizing lensing}

Here we provide a brief review of a technique to directly constrain the gravitational
lensing potential from the CMB power spectra introduced in \cite{Motloch:2016zsl,Motloch:2017rlk}.

\begin{figure}[b]
\center
\includegraphics[width = 0.49 \textwidth]{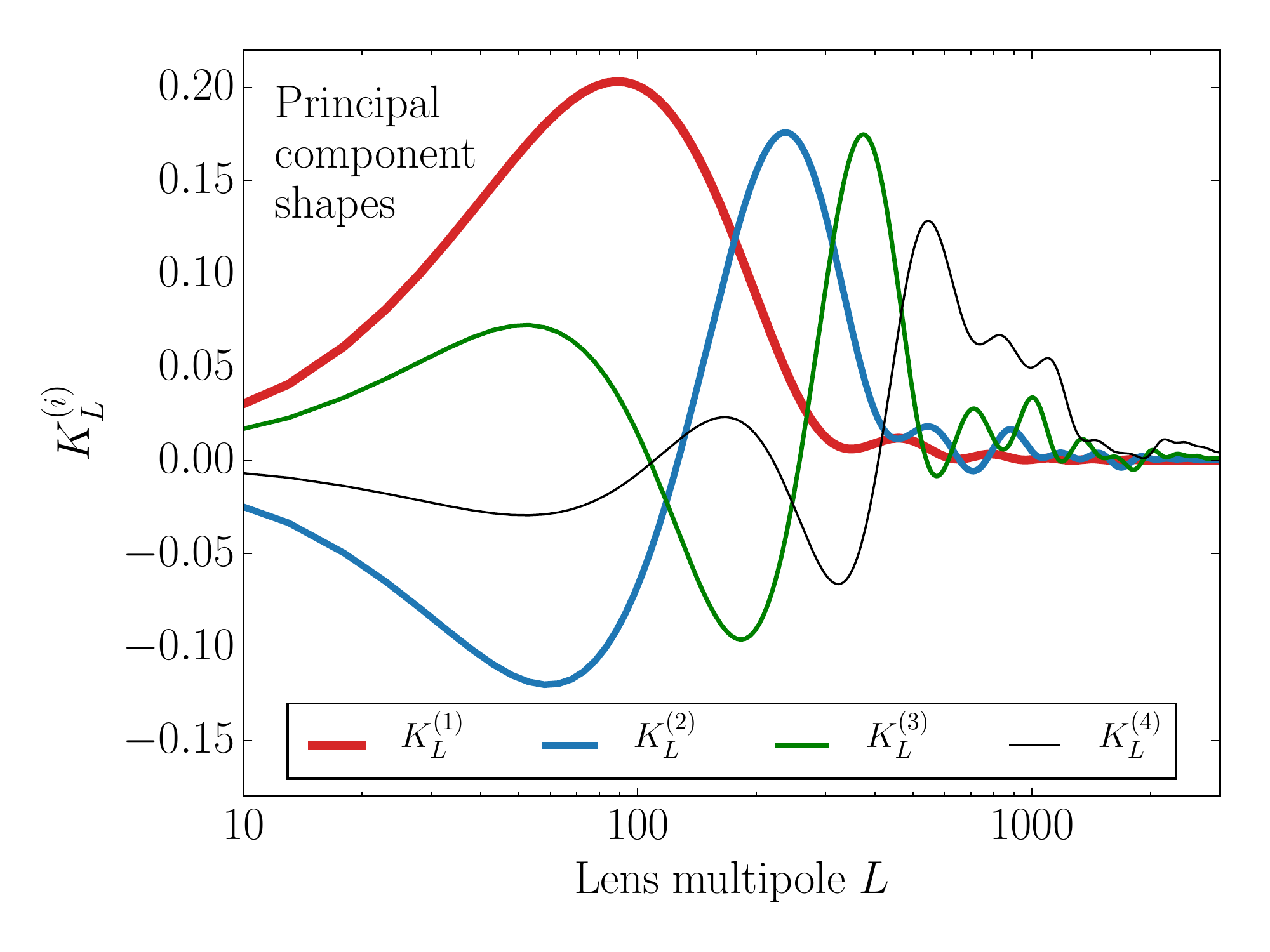}
\caption{Lensing principal components  $K_L^{(i)}$ used in this work.
}
\label{fig:pcs} 
\end{figure}

\begin{figure}[b]
\center
\includegraphics[width = 0.49 \textwidth]{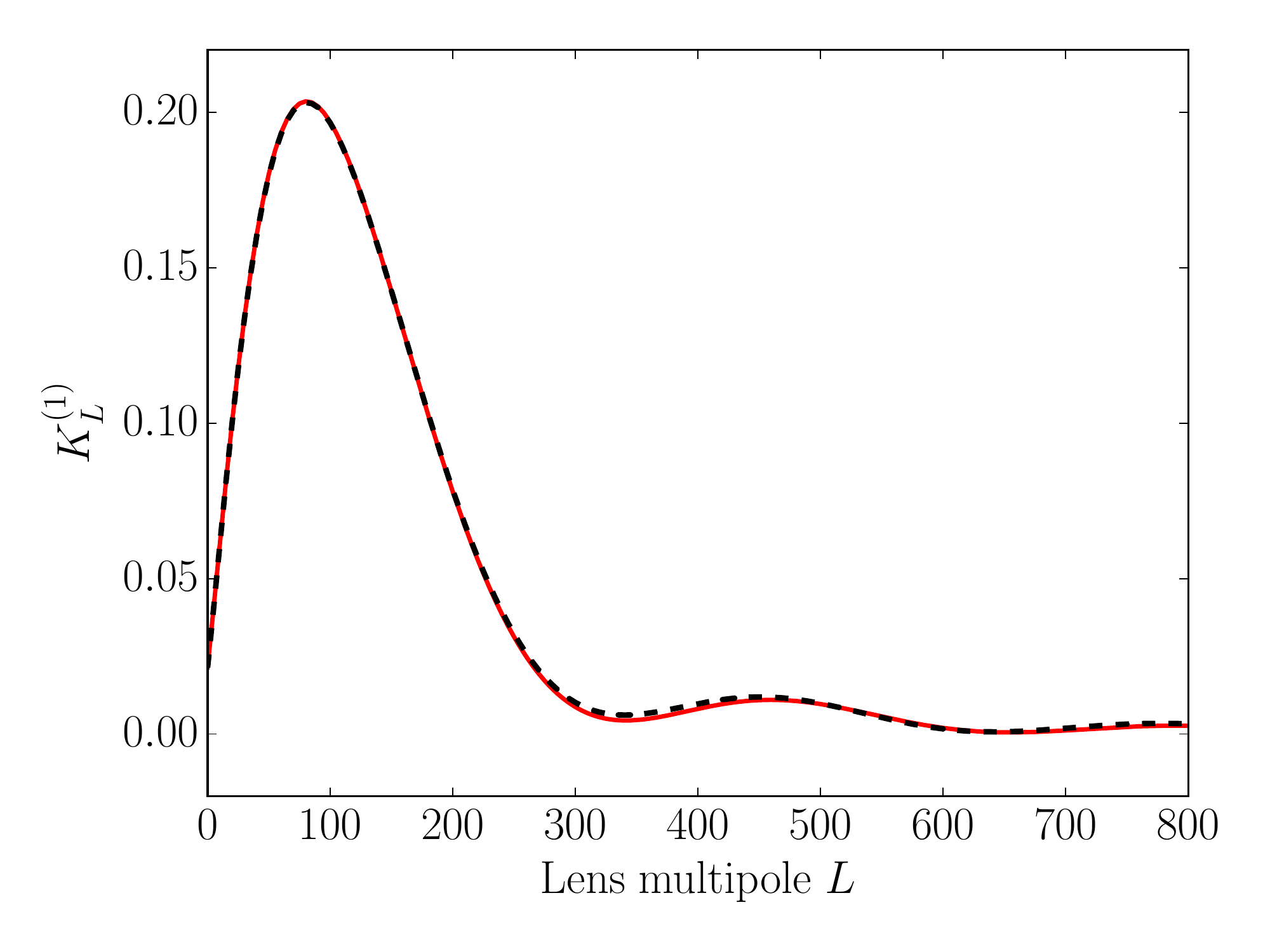}
\caption{Comparison of the shapes of the best measured principal component
$K_L^{(1)}$ as determined from the Planck TT (black dashed) and SPT TEEE (red) data.
They are very close, as is the one determined from SPT TT (not shown).
}
\label{fig:pcs_comparison} 
\end{figure}

The gravitational lensing potential power spectrum is parameterized in terms of 
$N_\mathrm{pc}$ effective parameters $\Theta^{(i)}$ which determine arbitrary
variations around a fixed fiducial power spectrum $C_{L, \mathrm{fid}}^\PP$ as 
\be
\label{definition}
	C_L^\PP = C_{L, \mathrm{fid}}^\PP \exp\(\sum_{i = 1}^{N_\mathrm{pc}} K^{(i)}_L\, \Theta^{(i)} \) .
\ee
In this setup, constraining $\Theta^{(i)}$ from the data corresponds directly to
constraining the gravitational lensing potential. This should be contrasted with the
common approach of introducing a phenomenological parameter $A_L$ which multiplies
$C_L^{\phi\phi}$ at each point in the model space and cannot be so interpreted once model
parameters are marginalized over.

We choose the same fiducial model and
$K_L^{(i)}$ as in \cite{Motloch:2018pjy} to allow easier comparison to those results.  More details about the fiducial model and values
of cosmological parameters are given in \S\,\ref{sec:simulations}. These $K_L^{(i)}$ are
chosen such that $\Theta^{(i)}$ correspond to $N_\mathrm{pc}$ principal components (PCs)
of the gravitational lensing potential best measured by \verb|Planck TT|, as determined
using a Fisher matrix construction \cite{Motloch:2018pjy}. The resultant eigenmodes
$K_L^{(i)}$ are shown in Fig.~\ref{fig:pcs}.  We retain $N_\mathrm{pc}=4$ PCs in order to
fully characterize all sources of lensing information \cite{Motloch:2018pjy}.
Accommodating the PCs to the SPT covariance is not necessary, as the shapes $K_L^{(1)}$
that would correspond to the lensing modes best constrained by the \verb|SPT TT| and
\verb|SPT TEEE| likelihoods are very close to the one derived from the \verb|Planck TT|
likelihood (Fig.~\ref{fig:pcs_comparison}). Additionally, the other principal components
in SPT have variance at least $\sim 100$ times larger than the leading PC and the data are
thus unable to constrain them strongly. Throughout this work we consistently use a single
set of $K_L^{(i)}$, given by \verb|Planck TT|.

In models beyond $\Lambda$CDM, changes in the integrated Sachs-Wolfe (ISW) effect would
typically affect data on the largest scales. In this work we are interested only in
lensing-like effects and leave the ISW contribution at its $\Lambda$CDM value.

\subsection{Markov Chain Monte Carlo sampling}

To sample the posterior probability in the various parameter spaces we use the MCMC code CosmoMC\footnote{https://github.com/cmbant/CosmoMC}
\cite{Lewis:2002ah}.  Each of our chains has a sufficient number of samples such that the
Gelman-Rubin statistic $R-1$ \cite{Gelman:1992zz} falls below 0.01.

We choose flat tophat priors on $\Theta^{(i)}$. As $\Theta^{(1)}$ is the variable in which
we will evaluate the tensions between data sets, we choose uninformative prior on it. For
the remaining three $\Theta^{(i)}$, that {allow freedom in the shape of the gravitational
lensing potential}, we limit their variation such that all $C_L^\PP$ are
within a factor of 1.5 of $C_{L,\rm fid}^\PP$. These weak priors are meant to eliminate
cases that would be in conflict with other measurements of large scale structure or imply
unphysically large amplitude high frequency features in $C_L^\PP$.

In analyses that use \verb|Planck TT|, \verb|SPT TEEE| or \verb|SPT TT|, in
addition to these four lensing parameters we also vary the six $\Lambda$CDM parameters,
with flat uninformative priors. Unlike the standard analysis, {which we also conduct for comparison},
 these only affect the
unlensed power spectra and their changes do not in any way affect the gravitational
lensing potential that is fully determined by $\Theta^{(i)}$ \eqref{definition}.

We use default foreground and nuisance parameters and their priors in all the
likelihoods.

\section{{Lensing covariance effects in SPT TEEE data}}
\label{sec:lic_in_spt_teee}

As we will see shortly, gravitational lensing measurements from \verb|SPT TEEE|
are so constraining that the lensing-induced covariance terms have to be included.
In this section we comment on possible ways  to account for this covariance in the
likelihood and what cosmological parameters are affected in the standard $\Lambda$CDM model and in its parameterized
lensing extension.

\subsection{Super-sample covariance}
\label{sec:ssc}

To obtain the data covariance matrix, the SPT collaboration used simulations 
based on Gaussian realizations of lensed CMB power spectra instead of actually lensing the
simulated CMB data. As a consequence, in this approach the lensing-induced covariance
terms are missing from their covariance matrix.

Instead of explicitly including the SSC term in the covariance matrix, the SPT
collaboration introduced a new parameter $\bar \kappa$ into the \verb|SPT TEEE|
likelihood.
{The parameter $\bar \kappa$ quantifies  the unknown value of the mean lensing convergence in the survey
which shifts the power spectra according to 
\begin{equation}
C_\ell^{XY}(p_\mu,\bar \kappa) = C_\ell^{XY}(p_\mu) + \frac{\partial \ell^2 C_\ell^{XY}}{\partial \ln \ell} \frac{\bar\kappa}{\ell^2},
\label{SSCparameter}
\end{equation}
where $p_\mu$ are the cosmological parameters}.\footnote{This technique was introduced in \cite{Manzotti:2014wca} but note
that $\bar \kappa \rightarrow -\bar\kappa$ in their Eq.~(32). } 
We find that including  super-sample
lensing as an additional covariance  by adding \eqref{ssc}  or through the additional parameter $\bar \kappa$ leads to
identical results.   Since the measurement of $\bar\kappa$ can be useful when comparing to other
data sets, as we show below, from this point forward we adopt it in our analysis.
When considering the SSC effect we also include the ILC covariance in the analysis and vice versa, but the 
{results} are not sensitive to its inclusion.

\begin{figure}
\center
\includegraphics[width = 0.49 \textwidth]{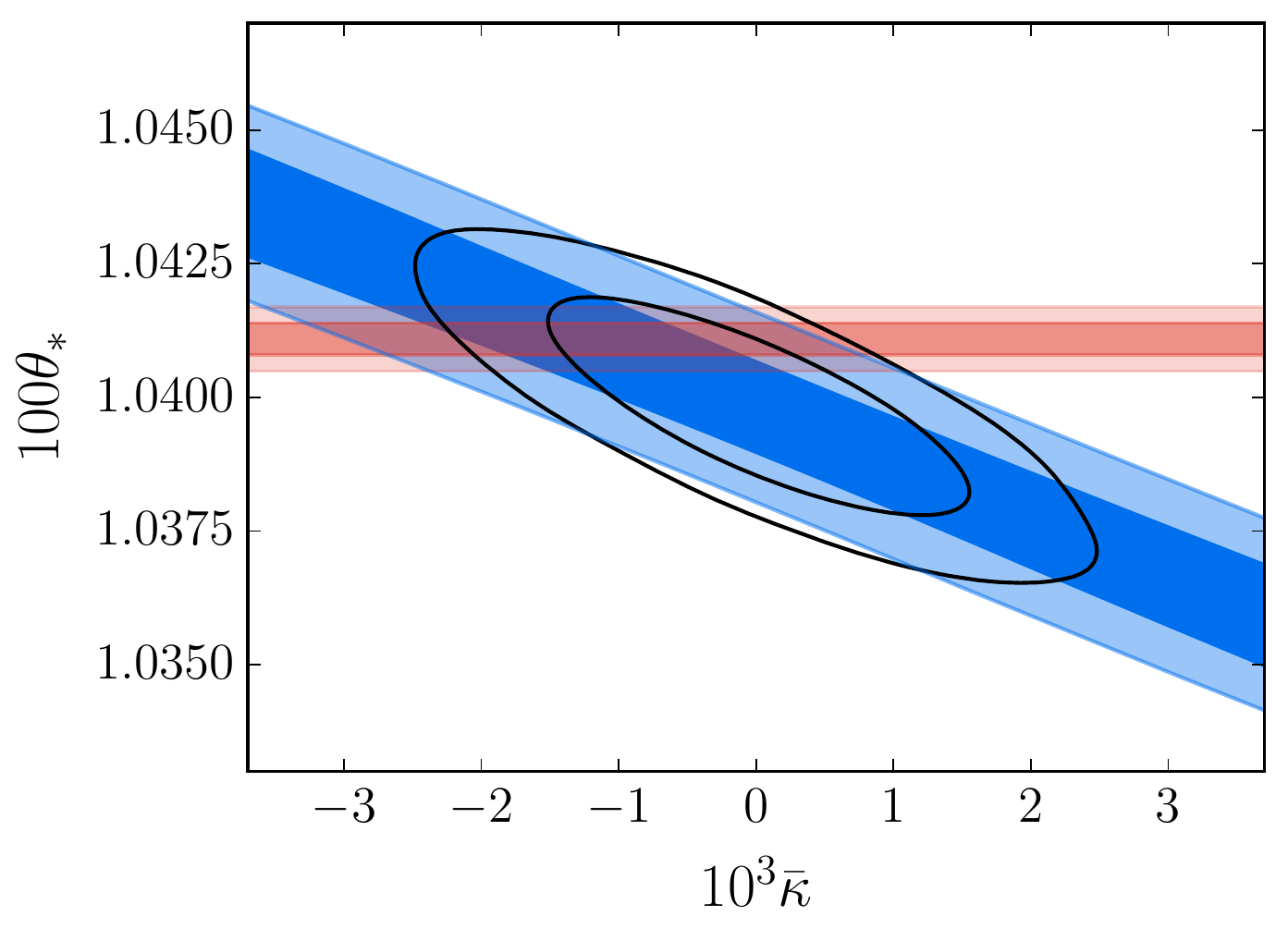}
\cprotect\caption{Constraints on $\theta_*$ and $\bar \kappa$ from \verb|SPT TEEE| when no
prior on $\bar \kappa$ is chosen (blue solid) and with the $\sigma_{\bar \kappa} =
10^{-3}$ prior (black lines). In red we show
constraints on $\theta_*$ from Planck 2018 temperature and polarization
cosmological parameter constraints \cite{Aghanim:2018eyx}. Results assume $\Lambda$CDM and
display 68\% and 95\% confidence limits. }
\label{fig:spt_kappa}
\end{figure}

Due to the strong degeneracy
with $\theta_*$ (see Fig.~\ref{fig:spt_kappa}), {the parameter $\bar \kappa$} is only
very poorly constrained by the SPT data itself and is limited by a Gaussian prior with
width $\sigma_{\bar\kappa} = 1.0 \times 10^{-3}$, 
reflecting the expected fluctuations of the super-sample lenses. The size of the prior
was chosen by the SPT collaboration according to \eqref{sigma_kappa_2}; using our fiducial
cosmology and SPTpol-like rectangular window we obtain a similar value. With $\bar\kappa$
prior, $\theta_*$ can be constrained, as evident from Fig.~\ref{fig:spt_kappa}.
The cosmological parameters other than $\theta_*$ are not significantly
affected by the super-sample lensing effect.

Using \verb|SPT TEEE| and assuming $\Lambda$CDM, the prior uncertainty on $\bar \kappa$ limits
the measurement of the angular scale of the angular peaks {to 
$100\theta_* = 1.03982 \pm 0.00135$.}\footnote{In \S\,\ref{sec:mcmc_results} we found that variance of
$\bar\kappa$ calculated by \eqref{sigma_kappa_2} appears to be underestimated by $\sim$
5\%. Even if this is  the case, the uncertainty on $\theta_*$ from {\tt SPT TEEE} would
grow only marginally, to $1.38 \times 10^{-5}$.   }  
For the 2018 Planck temperature and polarization data \cite{Aghanim:2018eyx}, the
impact of SSC is negligible which allows an extremely precise measurement of $\theta_*$:
$100\theta_* = 1.04109  \pm 0.00030$, with a difference in mean from  the \verb|SPT TEEE|
measurement that is {in good agreement at $0.9\sigma$.}  If SSC were ignored in the \verb|SPT TEEE| analysis, constraints on the
angular scale of the acoustic peaks would be too optimistic, $100\theta_* = 1.03985 \pm
0.00085$, {leading to an overly significant $1.4\sigma$ difference in means}.

The benefit of considering $\bar\kappa$ as a parameter is that when combined with the
Planck 2018 measurement of $\theta_*$, a more precise measurement of its value in the SPT field
can be extracted and compared to other measurements of lensing.
The Planck measurement breaks the $\bar \kappa - \theta_*$ degeneracy
in the \verb|SPT TEEE| without any need for the $\sigma_{\bar\kappa}$ prior.
This approach allows us to determine $\bar \kappa$ in the SPTpol field and leads to
\be
	\bar \kappa_\mathrm{SPTpol} = (-1.3 \pm 0.9) \times 10^{-3} .
\ee
The mean in the field is consistent with the expected root mean squared (rms)
$\sigma_{\bar\kappa}$ of the $\Lambda$CDM model and the errors {approach} the intrinsic
sensitivity of the \verb|SPT TEEE|  data to a fractional shift in angular scale in
the absence of the $\bar\kappa-\theta_*$ degeneracy, i.e.\ {$\sigma_{\theta_*}/\theta_*\approx  0.8 \times 10^{-3}$. }
 Combined they 
show a mildly significant indication of an  underdensity in the SPTpol footprint.   

In principle, this mild preference can be tested against other measurements of lensing, for example the Planck lensing map \cite{Ade:2015zua}.     However, the Planck lensing map is noisy and
band limited to $L \ge 8$, which removes part of the super-sample lensing signal. The quantity
\be
	\bar \kappa_\mathrm{SPTpol}^\mathrm{est} = \frac{1}{A} \int \mathrm{d}\hat n\, w(\hat n) \hat \kappa(\hat
	n) ,
\ee
where $\hat \kappa$ is the Planck lensing map, is an unbiased estimator of $\bar
\kappa_\mathrm{SPTpol}$ with variance
\be
	\sigma_{\bar \kappa_\mathrm{SPTpol}}^2 = 
	\frac{1}{A^2}\sum_{LM} \left|w_{LM}\right|^2 \frac{L^2(L+1)^2}{4}
	\chi_L^{\phi\phi} ,
\ee
where
\ba
	\chi_L^{\phi\phi} &=& 
	\begin{cases}
	 C_L^\PP, & L < 8\nonumber\\
	 N_L^\PP, & L \ge 8
	 \end{cases} 
	  ,
\ea
with $N_L^\PP$ being the noise power
 in the Planck lensing map. The $L < 8$ terms account
for the missing large scale lensing modes, while the $L \ge 8$ terms include the
uncertainty due to the noise in the Planck lensing map; we assume the noise is uncorrelated with the lensing
signal. Using our rectangular window function as a proxy for the real SPTpol mask, we
obtain
\be
	\bar \kappa_\mathrm{SPTpol}^\mathrm{est} = \(-0.7 \pm 1.2\) \times 10^{-3} ,
\ee
result consistent with the \verb|SPT TEEE| measurements.

We see from \eqref{sigma_kappa_2} that the expectation for the rms $\bar\kappa$, $\sigma_{\bar \kappa}$, depends on the
gravitational lensing potential and the SSC amplitude should in principle be evaluated in
each point in the parameter space.  
However,
in the $\Lambda$CDM model the shape and amplitude of lensing is sufficiently well constrained and consistent between measurements that parameter
uncertainties provide only small changes in its value compared with the measurement errors.  
We shall see that the same is not true of the parameterized lensing extension to $\Lambda$CDM and so we choose not
to repeat the SSC analysis of this section for this model  until such tensions are resolved.
Likewise, although as we shall see ILC can also be treated with auxiliary
parameters, because of these tensions we do not conduct such an analysis in this
work.

\subsection{Intra-sample lensing covariance}

As pointed out in the previous section, the \verb|SPT TEEE| covariance is based on
simulations without actual gravitational lensing and is thus missing the ILC term.
To include the ILC, we add $\Cov^{XY,WZ}_{(\ILC)bb'}$ calculated according to \eqref{ilc}
on top of the covariance matrix provided by the SPT collaboration in the \verb|SPT TEEE|
likelihood.  To calculate the ILC term here, we use the $\mathcal{U}^{XY}_{b\ell}$ that
are also provided in the \verb|SPT TEEE| likelihood.

In $\Lambda$CDM, we find that the main effect of adding the
intra-sample lensing covariance to \verb|SPT TEEE| is a degradation of the $\Omega_c h^2$
constraints by $\sim 7\%$. This is because part
of the information on $\Omega_c h^2$ comes from the smoothing of the peaks due to lensing and hence
quantifying the errors on lensing information is important for its determination.
Correspondingly, the constraints on the Hubble constant change from 
$H_0=\(70.8 \pm 2.1\)\,\mathrm{km/s/Mpc}$ 
to 
$\(70.4 \pm 2.3\)\,\mathrm{km/s/Mpc}$.

\begin{figure}
\center
\includegraphics[width = 0.49 \textwidth]{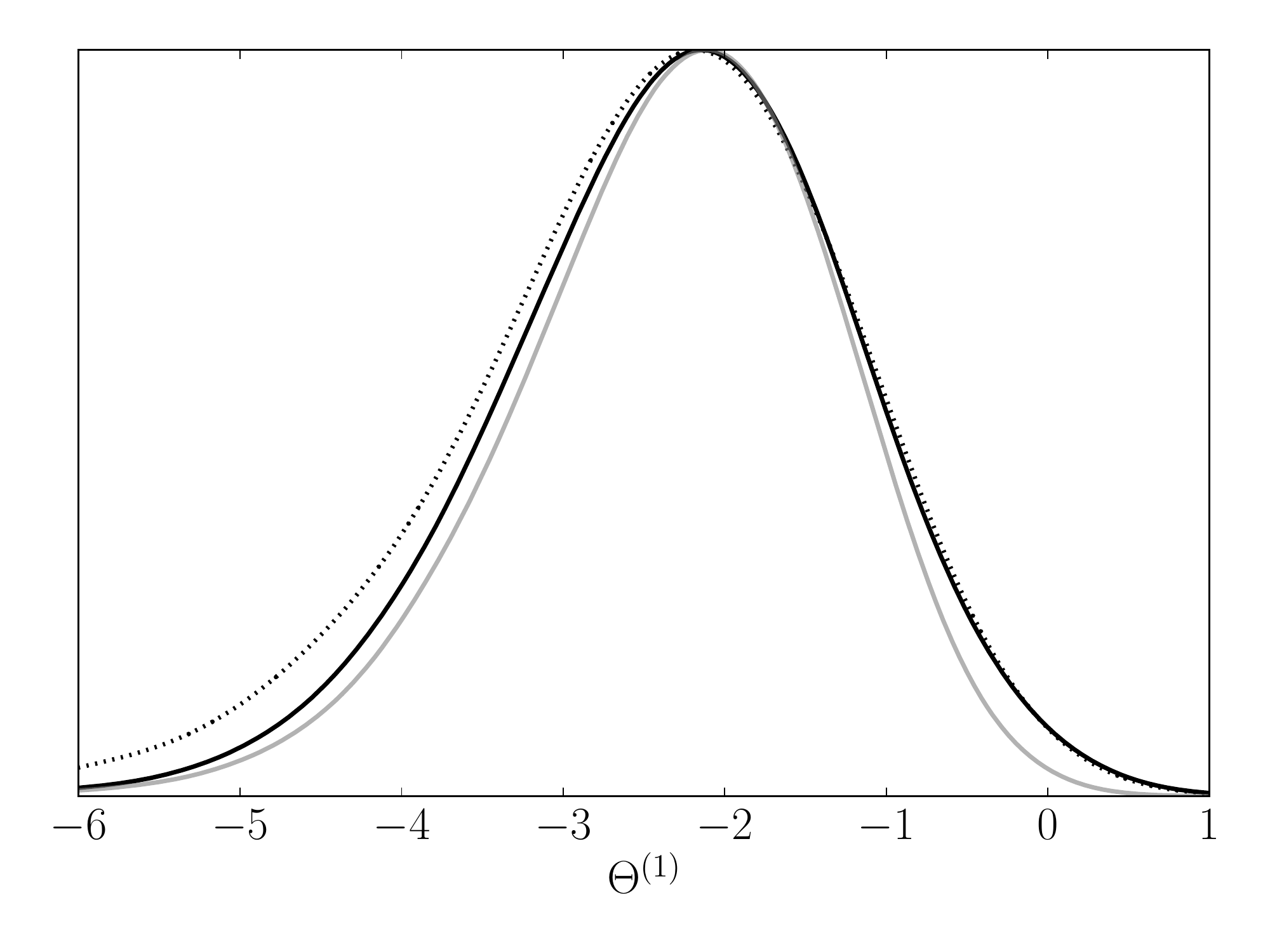}
\cprotect\caption{Comparison of $\Theta^{(1)}$ posterior probability distributions from
\verb|SPT TEEE| for various models of the ILC effect: original likelihood without ILC
(gray), constant ILC evaluated at the fiducial model (dotted) and $\Theta^{(1)}$-dependent
ILC (black solid).}
\label{fig:spt_teee_constraints}
\end{figure}

This impact on lensing information of the ILC can be best understood within the context of the model where
the gravitational lensing is separately parameterized in terms of the lensing PCs $\Theta^{(i)}$. In
Fig.~\ref{fig:spt_teee_constraints} we compare constraints on the best measured PC
$\Theta^{(1)}$ before and after addition of the ILC effect into the covariance.
As expected, adding ILC degrades the constraints, as we are
effectively adding uncertainty related to the unknown lens fluctuations.  
In $\Lambda$CDM, gravitational lensing information is mainly used to constrain $\Omega_c
h^2$, which explains the observed effect. 
In extensions that change the low-redshift physics, such as by allowing the mass of the
neutrinos $\sum m_\nu$ or equation of state of the dark energy parameter $w$ to vary,
 the ILC
most affects constraints on combinations of the $\Lambda$CDM and extension parameters
that are limited by the lensing information, see \cite{Motloch:2017rlk} for a related
discussion.

As evident from \eqref{ilc}, the magnitude of the ILC effect grows with increasing
gravitational lensing power. 
For \verb|SPT TEEE| this dependence must be considered, because the
constraints on gravitational lensing from \verb|SPT TEEE|, shown in
Fig.~\ref{fig:spt_teee_constraints}, are still rather weak.
To partially account for this effect, we explicitly model
the dependence of $\Cov^{XY,WZ}_{(
\ILC)}$ on the dominant lensing component $\Theta^{(1)}$: we
evaluate \eqref{ilc} for a representative set of gravitational lensing potentials
corresponding to $\Theta^{(1)}$ in the range constrained by \verb|SPT TEEE| (see
Fig.~\ref{fig:spt_teee_constraints}).  We then interpolate to get a smooth dependence on
$\Theta^{(1)}$ of the ILC contribution to the covariance matrix and reevaluate the
covariance matrix in each point in the parameter space.

The main effect of considering the $\Theta^{(1)}$-dependence in ILC, as opposed to using
ILC evaluated for the fiducial lensing potential $C^\PP_{\mathrm{fid},L}$, is a suppressed
probability of low values of $\Theta^{(1)}$ (Fig~\ref{fig:spt_teee_constraints}). This can
be easily understood, as lower values of $\Theta^{(1)}$ correspond to smaller lensing
power, leading to smaller amplitude of the ILC effect; the constraints at low
$\Theta^{(1)}$ then effectively approach the case without ILC. From this point forward
we use this model of ILC for  the \verb|SPT TEEE| analysis.

The addition of lensing-induced covariance to
either of the \verb|PP| likelihoods is not necessary, as the lens variance is
already included in the Gaussian terms of their covariance matrices. 
Furthermore, due to the larger sky coverage
and correspondingly smaller lens variance  (recall the $A^{-1}$ factor in \eqref{ilc})
{and} the fact that $C^{TT}_\ell$ is less sensitive to lensing effects than the
polarization power spectra, ILC modifications are not necessary for the \verb|Planck TT| and \verb|SPT TT|  likelihoods. 

As a rule of thumb, the ILC effect has to be included when the $\Theta^{(1)}$ constraint
from the given data set  approaches
\be
\label{cv_sigma_Th1}
	\sigma^A_{\Theta^{(1)}} \approx 0.048\sqrt{\frac{4\pi}{A}},
\ee
which is the limiting error due to the sample variance of
 $\Theta^{(1)}$ in a given patch of the sky.  
 
Finally, given that future experiments will have to include the ILC effect
into their analysis but will also provide
 much tighter constraints on lensing effects as they approach this sample variance limit, we conclude
 with a simple approach to incorporating the ILC effect for the purpose of
 cosmological parameter estimation.  In this context  $\Cov^{XY,WZ}_{(\mathrm{ILC})bb'}$
 can be considered constant and given by the
 best fit parameters.   The addition of
ILC to the analysis can then be done in a way that parallels the treatment of SSC through $\bar
\kappa$ by considering $\bar\Theta^{(i)}$ as parameters that describe a local
fluctuation in the lens power spectra within the survey footprint,
affecting the CMB power spectra as
\begin{equation}
C_\ell^{XY}(p_\mu,\bar\Theta^{(i)} ) = C_\ell^{XY}(p_\mu) +\sum_i \frac{\partial C_\ell^{XY}}{\partial\Theta^{(i)}}
\bar\Theta^{(i)} .
\end{equation}
Here $p_\mu$ are the cosmological parameters of the model and the sum is over a sufficient number of principal components, {either constructed from a Fisher matrix for the experiment as we have done here for Planck}  or by empirically discovering them from simulations
 as described in Appendix  \ref{sec:app_alternative_method}.
To account for the effect of ILC on cosmological parameter estimation,
one then marginalizes over $\bar\Theta^{(i)}$ given a theoretical prior on the amplitude of
the lens fluctuations within the window.
Note however that  this procedure assumes that there is a consistent cosmological model that describes
all lensing effects in all datasets, which is not currently the case, as we shall see next.

\section{SPT-Planck lensing tensions}
\label{sec:tensions}

\begin{figure}
\center
\includegraphics[width = 0.49 \textwidth]{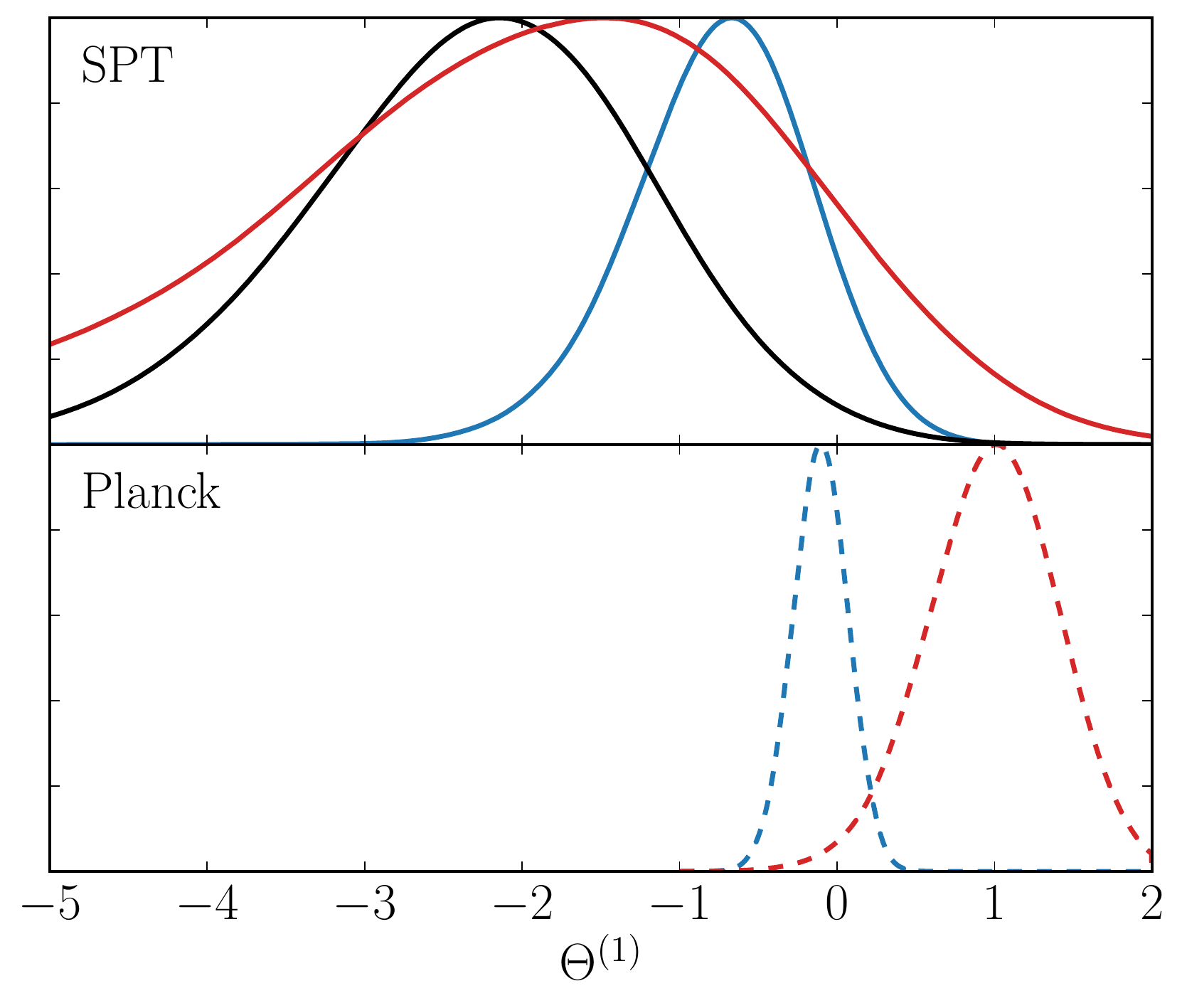}
\cprotect\caption{ Constraints on $\Theta^{(1)}$ from SPT (top) and Planck (bottom, dashed): in blue, lensing constraints from lensing reconstruction (\verb|SPT PP| and \verb|Planck PP|); in
red, from temperature power spectra (\verb|SPT TT| and \verb|Planck TT|); in black,
from \verb|SPT TEEE| with the ILC effect included (black solid).}
\label{fig:all_constraints}
\end{figure}

{In this section we compare lensing constraints from the various SPT and Planck data sets using
the  techniques developed in the previous sections.}
In Fig.~\ref{fig:all_constraints} we compare lensing constraints from all five
data sets investigated in this work, including the \verb|SPT TEEE| data set with a
$\Theta^{(1)}$-dependent covariance matrix. 

All lensing constraints from SPT are mutually
consistent, while the Planck and SPT constraints seem to be systematically offset, with
SPT preferring lower values of the lensing potential (see
\cite{Henning:2017nuy,Simard:2017xtw} for related studies). 
This difference corresponds to sharper acoustic peaks in the SPT data compared to Planck data.
Sharper peaks cannot be caused by missing contributions from the super-sample lenses in the
smaller SPT sky area (see
Appendix~\ref{sec:app_sharpening}).

To quantify the significance of the tensions, we use a generalization to non-Gaussian
distributions of the commonly used ``shift in the means'' statistic; this generalization
is described in the Appendix~\ref{sec:app_significance}. In this work we assume all the
measurements are independent.

\begin{figure}
\center
\includegraphics[width = 0.49 \textwidth]{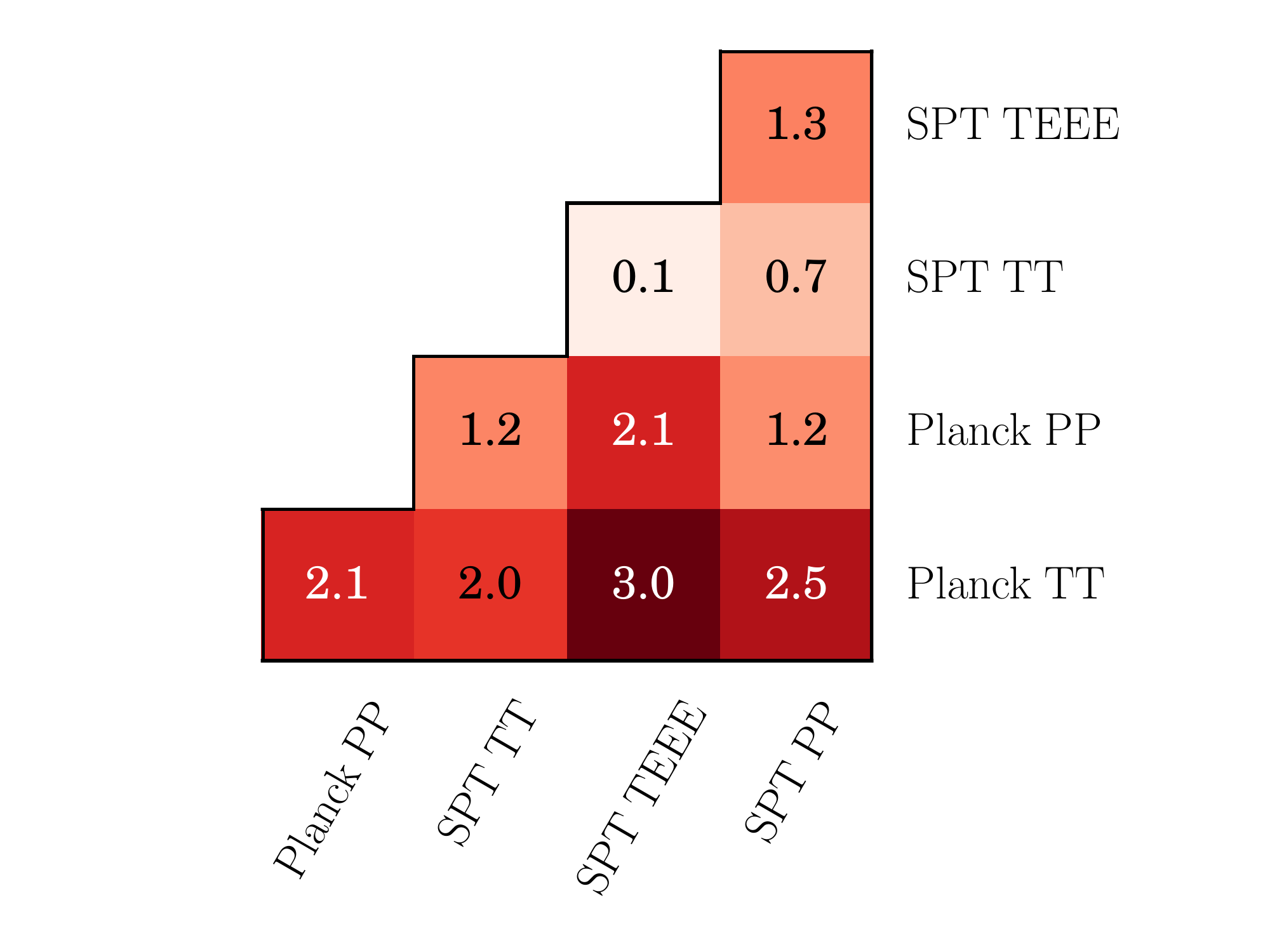}
\caption{Significance of the tensions between constraints on $\Theta^{(1)}$ from various
datasets. The numbers represent tensions in the units of Gaussian standard deviations
$\sigma$, so $1.0$ in this table corresponds to 31.7\% chance probability to exceed.}
\label{fig:tensions}
\end{figure}

Resulting tension significances
are listed in Fig.~\ref{fig:tensions}. The \verb|Planck TT| constraint
is in over 2$\sigma$ tension with all the other lensing constraints; its tension with
\verb|SPT TEEE| reaching 3.0$\sigma$ level. The constraint from this SPT likelihood is also in
a moderate {$2.1\sigma$} tension with the Planck lensing reconstruction constraint.

Had we not included the ILC effect into the covariance matrix, as was the case with previous analyses, 
the tensions between
\verb|SPT TEEE| and \verb|Planck TT| or \verb|Planck PP| would noticeably
increase {from 3.0$\sigma$} to 3.5$\sigma$ and  {from 2.1$\sigma$ to} 2.5$\sigma$ respectively. 
Including the dependence of
ILC on $\Theta^{(1)}$ is responsible for approximately 0.05$\sigma$ of the decrease.
Finally, let us point out that had we used the 2015 Planck lensing likelihood instead of
the 2018 update, the tension with \verb|SPT TEEE| would increase by additional
0.05$\sigma$.

Tension between \verb|SPT TEEE| and \verb|Planck TT| was investigated using
the standard $A_L$ parameter in \cite{Henning:2017nuy}. By comparing the SPTpol constraint
$A_L = 0.81 \pm 0.14$ with the Planck temperature constraints $A_L = 1.22 \pm 0.10$, {the difference in means is} in  2.4$\sigma$ tension\footnote{There is a typo in
\cite{Henning:2017nuy}, and the claimed 2.9$\sigma$ tension should be 2.4$\sigma$.}. This
is significantly less than the value 3.5$\sigma$ we find when not including the ILC effect
and clearly shows that comparison of $A_L$ is suboptimal, due to the fact that each $A_L$
scales the lens potential of a different cosmological model, i.e.\ those preferred by SPT
and Planck respectively.
We checked explicitly that when one considers the full seven-parameter
posterior of the $\Lambda$CDM+$A_L$ model, \verb|SPT TEEE| and \verb|Planck TT|
constraints on one particular linear combination of these seven parameters disagree at
the 3.5$\sigma$ level. The tension is thus in principle discoverable also in the standard
approach using $A_L$, but it is hidden in a combinations of parameters {and subject to interpretation
on parameter counting or the ``look elsewhere" effect} (see
\cite{Raveri:2018wln} for a related study).   Here we show that the tension is associated directly with 
the lensing effect on power spectra.  Moreover, $A_L$ will not be adequate  in the future, when the CMB power
spectra 
constrain more than just the amplitude of $C_L^\PP$.

With the exception of $\Theta^{(1)}$, constraints on all the other parameters, i.e.\ the six $\Lambda$CDM parameters as determined from the data through their effect on the
unlensed CMB, from \verb|Planck TT| and \verb|SPT TEEE| are mutually consistent; this is
in agreement with findings of \cite{Henning:2017nuy}.
For example, we find that after marginalizing over $\Theta^{(i)}$, the constraint
 on the Hubble constant
 becomes
  $H_0 = \(68.1 \pm 2.8\)\,\mathrm{km/s/Mpc}$
for   \verb|SPT TEEE| and $\(69.0 \pm
1.2\)\,\mathrm{km/s/Mpc}$ for
\verb|Planck TT|.

\begin{figure}
\center
\includegraphics[width = 0.49 \textwidth]{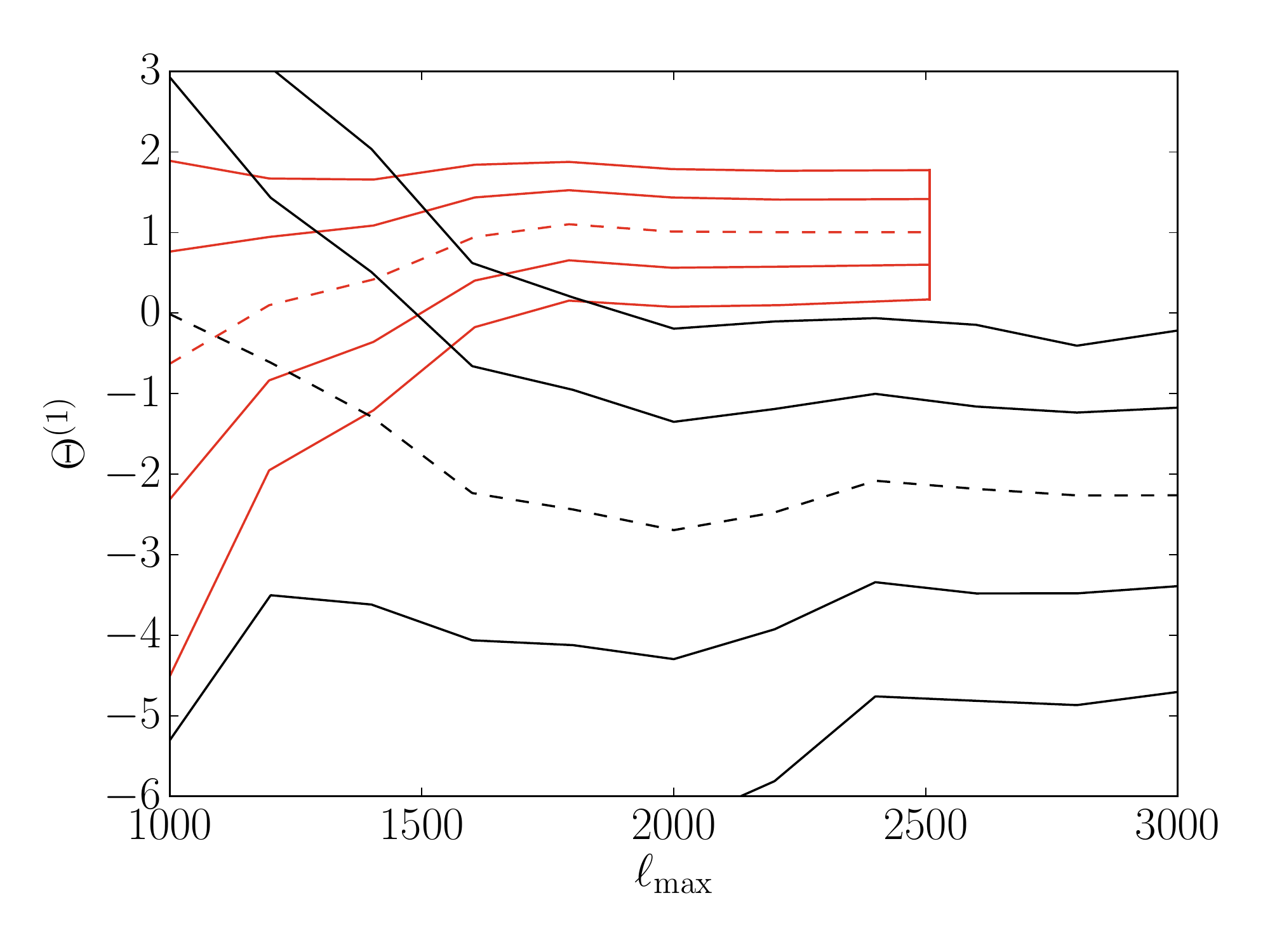}
\cprotect\caption{Constraints on $\Theta^{(1)}$ from
\verb|SPT TEEE| (black) and \verb|Planck TT| (red) likelihoods when only
part of the data up to the maximal multipole $\ell_\mathrm{max}$ is used. The
dashed lines show maximum likelihood values and the solid lines mark 68\% and 95\%
confidence intervals.} 
\label{fig:theta1_as_fn_of_lmax}
\end{figure}

To gain additional insight into the 3$\sigma$ tension between \verb|SPT TEEE| and
\verb|Planck TT|, in Fig.~\ref{fig:theta1_as_fn_of_lmax} we compare the constraints on
$\Theta^{(1)}$ from these likelihoods as a function of the maximal $\ell_\mathrm{max}$
considered in the analysis. The tension is generated in the $\ell$ range between 1000 and
2000, with the two likelihoods pulling in the opposite directions. While the lensing constraining
power of $\ell = 2000-3000$ in \verb|SPT TEEE| is comparable to that of the $\ell \le
2000$, there is no additional shift in $\Theta^{(1)}$.

\begin{figure*}
\center
\includegraphics[width = 0.49 \textwidth]{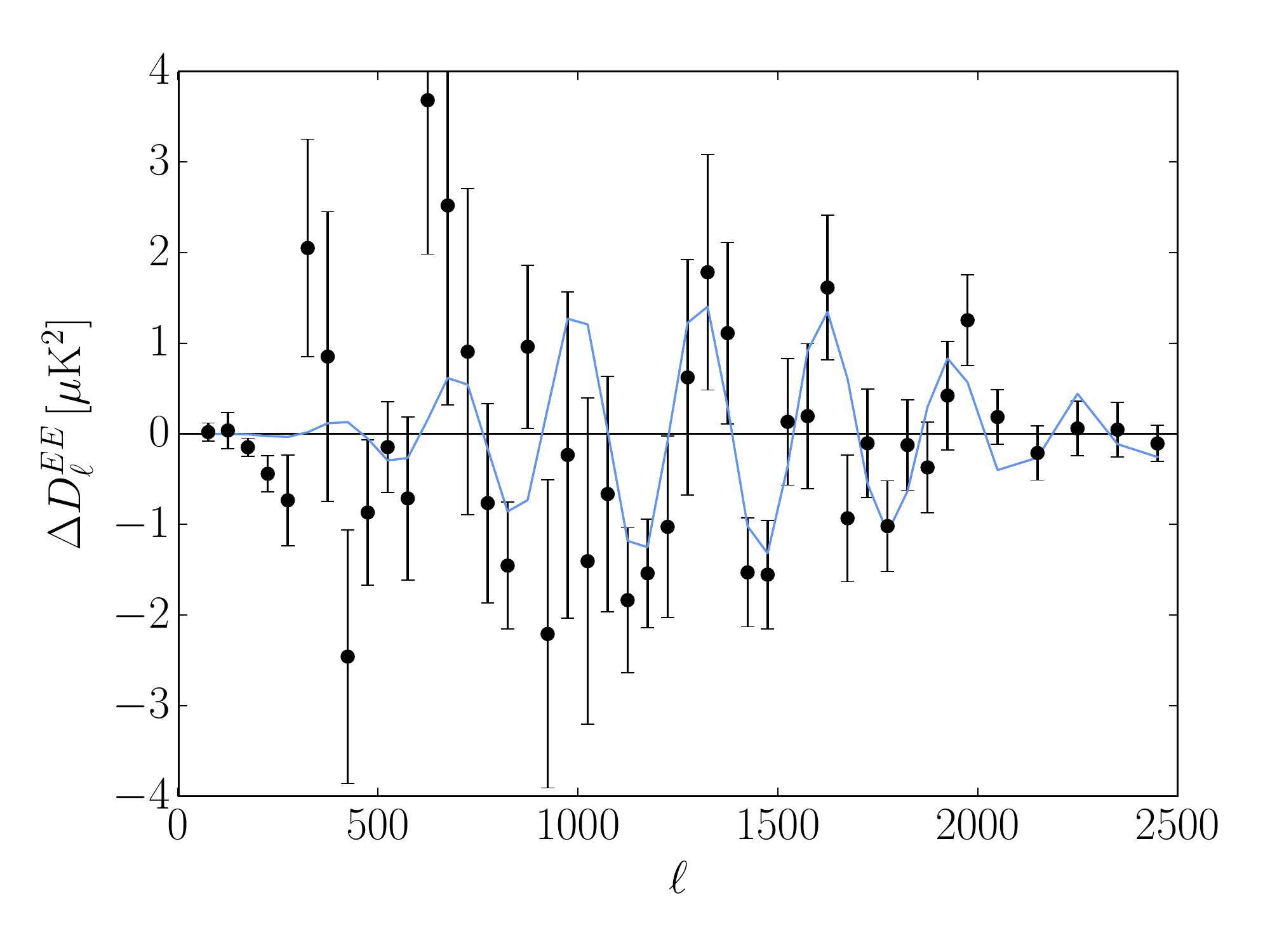}
\includegraphics[width = 0.49 \textwidth]{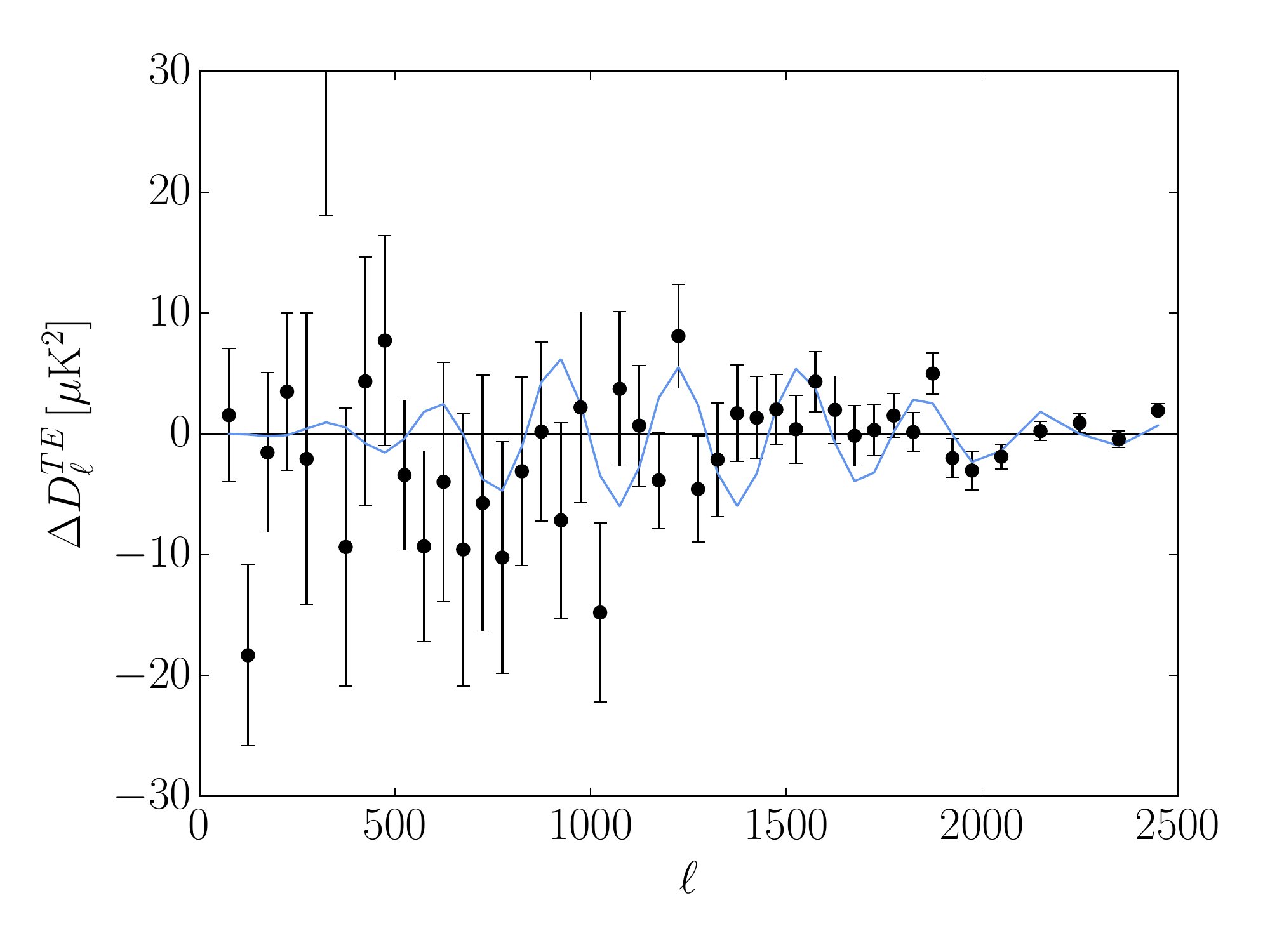}
\cprotect\caption{Points show the difference between the SPT measured values of
$D^{EE}_\ell$ (left) and $D^{TE}_\ell$ (right) and the best fit $\Lambda$CDM + 4
$\Theta^{(i)}$ model to the \verb|Planck TT|. For comparison, in blue we show $-7
\times \partial C^{XY}_\ell/\partial \Theta^{(1)}$, binned with the same binning scheme.}
\label{fig:residuals}
\end{figure*}

In  data space, the preference for high/low lensing shows as anomalously smooth/sharp
acoustic peaks, which allows for a clear illustration of the tension. The $D^{XY}_\ell =
{\ell(\ell+1)}C_\ell^{XY}/{2\pi}$ residuals between the SPT data and the best fit
$\Lambda$CDM + 4 $\Theta^{(i)}$ model to the \verb|Planck TT| data  are shown
in Figure~\ref{fig:residuals}.   The residuals in $C^{EE}_\ell$ exhibit distinct oscillations in
the $\ell$ range 1100-2200, with enhanced power at $EE$ peaks and reduced at troughs, consistent
with a deficit in lensing.
 Note that the $TT$ peaks are out of phase with $EE$ so that the smoothing in Planck $TT$ and sharpening
 in SPT $EE$ data occur at different multipoles, which makes this discrepancy difficult to explain with
  any physical mechanism.

\section{Discussion}
\label{sec:discussion}

In the first part of this work, we use simulated lensed CMB data to perform the first
investigation of covariances between and within $TT$, $TE$ and $EE$ power spectra measured
by experiments observing a small patch of the sky. We find that in general the
lensing-induced covariances are well described by a sum of the super-sample covariance,
parameterizing effects of lenses larger than the footprint, and intra-sample lensing
covariance, parameterizing effects of the smaller lenses. 

{As detailed in Appendix~\ref{sec:app_template_fitting},} the
amplitude of the ILC portion of the covariance is consistent with theoretical expectations whereas the SSC
portion  is about 5\% larger in our simulations than
theoretically expected.   For a typical analysis, this discrepancy does not have any {important consequence --
at most it would increase {the measurement errors on the angular extent of the sound horizon
$\theta_*$ by the same amount in cases when 
SSC limits such measurements,} {i.e for} small footprints.
 In  Appendix~\ref{sec:app_alternative_method}, {we confirm and refine these results, expanding on the
method of \cite{HarnoisDeraps:2011rb} which empirically extracts smooth features from a noisy estimate of
a correlation matrix, effectively decreasing the
numerical noise due to a limited number of simulations. In this work we use it to {extract}
 the SSC and ILC terms, ignoring terms close to the (sub)diagonals where the
Gaussian and window function effects dominate, but this method is applicable to any noisy estimate of a
covariance matrix.} 

Likewise, the method of treating SSC {by introducing} an auxiliary parameter 
in the window can be extended  to treat the {principal} modes of any such covariance matrix. For example, while we have omitted an analysis of  $C^{BB}_\ell$ here  due to complications from removing
the  $C^{EE}_\ell$ contamination {caused by} intermixing
due to the mask, this technique can be straightforwardly implemented by modeling lensing
covariance as a set of
extra parameters to marginalize over in the model for $C^{BB}_\ell$.  However, such an
analysis requires  a {known} prior expectation for the distribution 
of these parameters, which in practice requires first a resolution of lensing tensions in the current data. 

Finally extra sharpening of the acoustic peaks related to missing peak smoothing by the
super-sample lenses {is not detectable in our suite of simulations} and is constrained to be negligible even for cosmic
variance limited experiments in the investigated range of survey footprints, 150 deg${}^2$
-- 1000 deg${}^2$.

{In the second part of this work, we apply our lensing induced covariance analysis to South Pole Telescope
measurements. }
We find that the South Pole Telescope polarization constraints \cite{Henning:2017nuy} have
reached the levels of precision where the lensing-induced covariance terms have to be
included in the analysis.  Starting with the current generation of the CMB experiments,
these effects will thus have to become a standard part of CMB data likelihoods. {We show
how the non-Gaussian ILC effect can be added in an analytic way to a covariance matrix based on
Gaussian CMB assumptions, for example from Gaussian realizations of the power spectra.}
 When the covariance matrix is calculated using lensed CMB
simulations, the lensing-induced covariance in automatically included. In this case, the
technique of Appendix~\ref{sec:app_alternative_method} can be used to diminish the
numerical noise.

Parameterizing the mean lensing convergence in the SPTpol field $\bar
\kappa_\mathrm{SPTpol}$ and modelling it explicitly, instead of including the super-sample
lensing into the SSC covariance matrix, leads to identical results  when
considering the SPTpol results alone -- a nearly 60\% increase
in the uncertainty on $\theta_*$.   When combined with Planck information on
$\theta_*$ {within the $\Lambda$CDM model}, it enables us to constrain this parameter
$\bar \kappa_\mathrm{SPTpol}$ from the data.
The \verb|SPT TEEE| data hint at underdensity in the SPTpol region, which is in agreement
with the convergence calculated directly from the Planck lensing map.
Adding ILC within $\Lambda$CDM leads to approximately 10\% increase of the error
bars of $\Omega_c h^2$, which is the parameter for which the gravitational lensing
information is the most important.

Using the technique from \cite{Motloch:2016zsl} and including both the SSC and ILC effects, 
we obtain direct constraints on
gravitational lensing for the various South Pole Telescope likelihoods and compare them
against the Planck satellite constraints \cite{Motloch:2018pjy}. Because of non-Gaussian
posteriors, we generalize the standard ``shift in the means'' statistic to determine
tensions between the individual data sets, see Appendix~\ref{sec:app_significance}. While
the various constraints from  SPT are mutually consistent, we confirm that the SPT data
sets prefer relatively low lensing power; the tension between
\verb|SPT TEEE| and \verb|Planck TT| or \verb|Planck PP|  is significant at $3.0\sigma$ or 2.1$\sigma$ respectively. Preference of SPT data for low lensing power was previously found in
analyses based on the scaling parameter $A_L$ \cite{Henning:2017nuy,Simard:2017xtw},
however we find that when lensing tension between \verb|SPT TEEE| and \verb|Planck TT| is
investigated using $A_L$, its significance is severely underestimated. Using the technique
from \cite{Motloch:2016zsl} reveals the full lensing tension, and is thus recommended for
comparing lensing constraints across various data sets. The inclusion of the ILC into the
\verb|SPT TEEE| likelihood strongly affects {the probabilities to exceed} the
observed tensions. 
Had we not
included it, the tensions between \verb|SPT TEEE| and Planck constraints would grow by
about 15\%, reaching 3.5$\sigma$ or 2.5$\sigma$ between \verb|SPT TEEE| and
\verb|Planck TT| or \verb|Planck PP| respectively.

Constraints on $C_L^\PP$ from ongoing and upcoming CMB experiments such as SPT 3G \cite{Benson:2014qhw}, 
Advanced ACT \cite{Henderson:2015nzj}, 
Simons Observatory \cite{Ade:2018sbj} and CMB-S4 \cite{Abazajian:2016yjj} are
expected to significantly improve the lensing constraints.  The techniques developed in this work should prove even
more important in quantifying and resolving these tensions in the future.

\acknowledgements{
We thank Tom Crawford, Jason Henning, Ue-Li Pen, Marco Raveri and Kimmy Wu for useful discussions.
This work was
supported by NASA ATP NNX15AK22G and the Kavli
Institute for Cosmological Physics at the University of Chicago through grant NSF
PHY-1125897 and an endowment from the Kavli Foundation and its founder Fred Kavli.  WH was
additionally supported by   U.S.~Dept.\ of Energy contract DE-FG02-13ER41958  and the Simons Foundation.  
We acknowledge use of the CAMB, Lenspix and CosmoMC software packages. 
This work was completed in part with
resources provided by the University of Chicago Research Computing Center. 
}

\appendix

\section{{Simulated vs.\ theoretical covariances}}
\label{sec:app_agreement}

In this Appendix we quantify the agreement between the simulated covariance matrices
$\Cov^{XY,WZ}_{bb'}$ and their theoretical expectations. In the first part of the
Appendix, we model the lensing-induced covariance as a sum of SSC and ILC terms with
undetermined amplitudes and determine these amplitudes by minimizing the
residuals {versus the simulations}. In the second part we present results based on an alternative quantification
approach, expanding an eigenmode decomposition idea from
\cite{HarnoisDeraps:2011rb}, {which empirically
isolates the SSC and ILC effects.}

\subsection{Template fitting}
\label{sec:app_template_fitting}

In Fig.~\ref{fig:fit_residuals} we show the residuals between the correlation matrix
${\hat R^{XY,WZ}_{bb'}}$ obtained from simulations for the rectangular window function and the
theoretical expectation used throughout the main text,
{$R^{XY,WZ}_{(\mathrm{theory})bb'}$}. The residuals are small
and appear noise-like, with the possible exception of $R^{TT,TT}$ at high $\ell$ {which show hints
of structure unrelated to the SSC or ILC template forms}. As
similar residuals do not appear for the other window functions considered here, {indicating that it may be an artifact of the limited number of simulations}, we do not
investigate this issue further.

\begin{figure*}
\center
\includegraphics[width = 0.49 \textwidth]{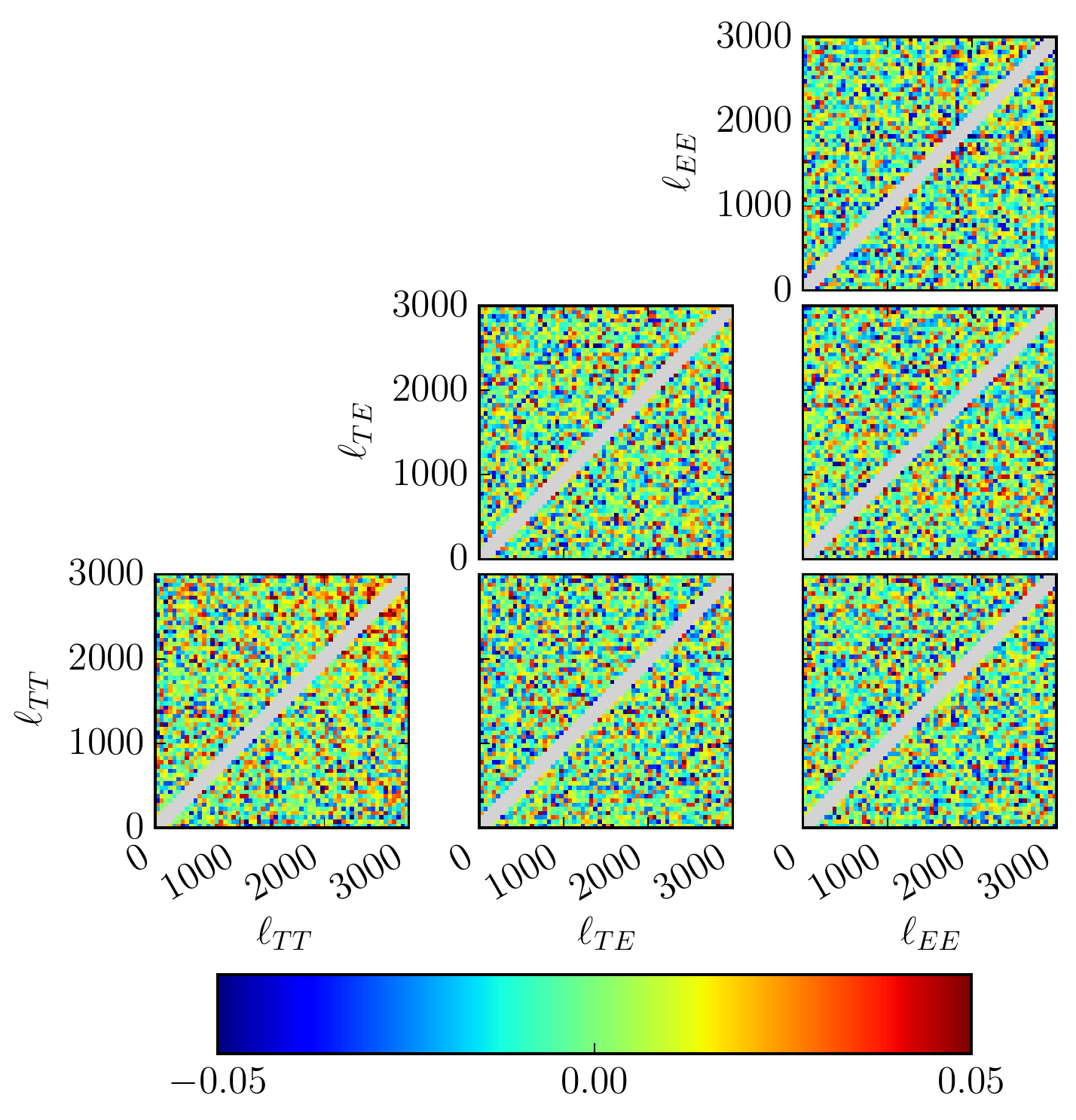}
\includegraphics[width = 0.49 \textwidth]{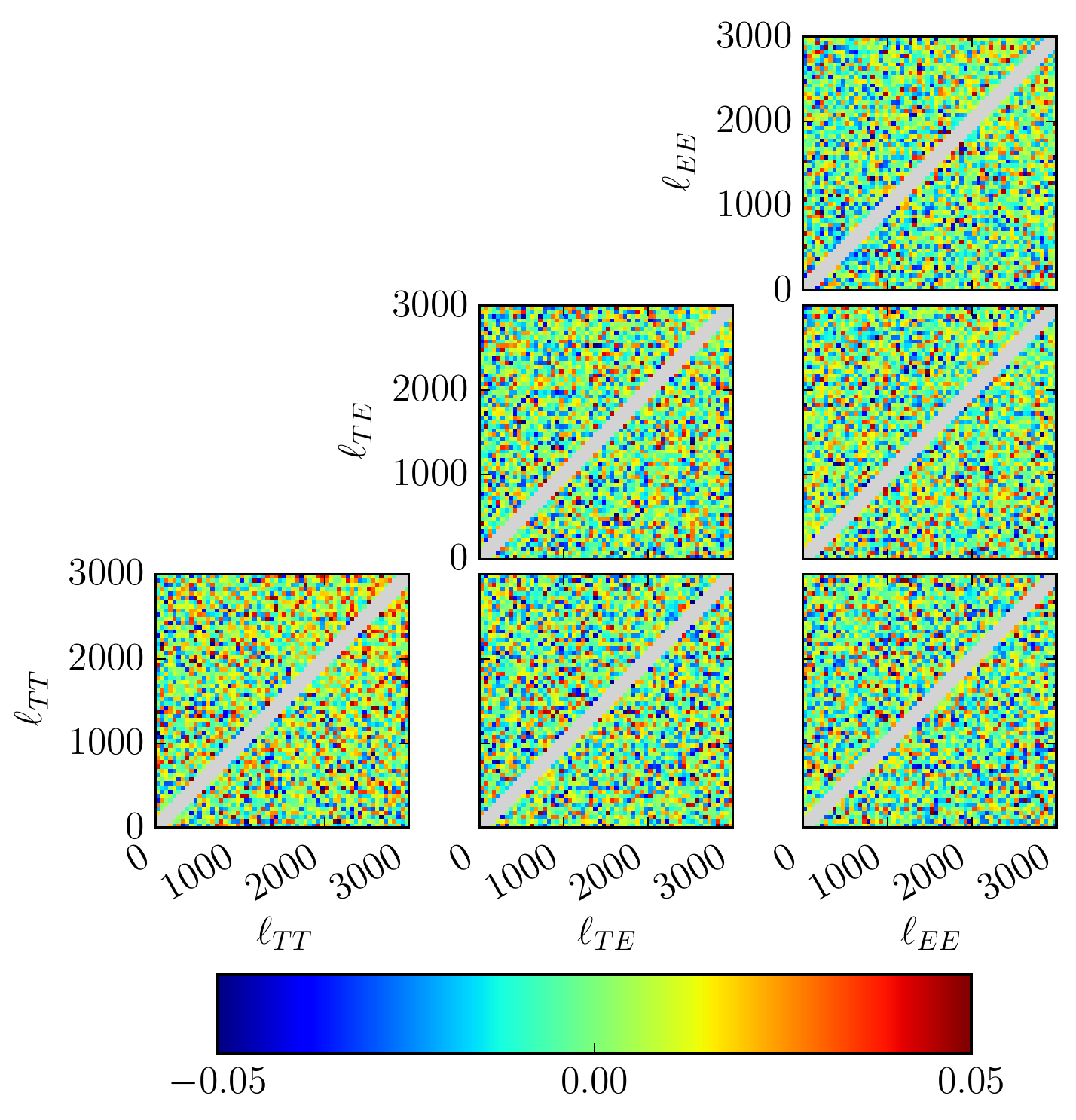}
\caption{
Left:
Residuals between the correlation matrix ${\hat R^{XY,WZ}_{bb'}}$ obtained from simulations
and the theoretical expectation for the lensing-induced correlation {with no rescaling of the SSC and ILC
effects.  Residuals are on a substantially smaller scale than the effects shown in 
Fig.~\ref{fig:mcmc_cov_matrix} and are mainly consistent with noise (see text).}
  The gray region hides elements dominated by Gaussian
covariance terms and the window function effects.
Right: Analogous residuals between the correlation matrix ${\hat R^{XY,WZ}_{bb'}}$ obtained from simulations
and its approximation through \eqref{ueli_approx}, again for the rectangular window function.
}
\label{fig:fit_residuals}
\end{figure*}

On the other hand, we can test the amplitude of the SSC and ILC effects considering their
form as given.  To do so, we look at the matrix elements away from the diagonals, where the
Gaussian terms and window function effects are negligible. We focus on bins $|b - b'| > 2$
and model the covariance there as
\be
\label{model_covariance}
	\Cov^{XY,WZ}_{(\mathrm{model})bb'} = 
	A_1 \Cov^{XY,WZ}_{(\SSC)bb'} +
	A_2 \Cov^{XY,WZ}_{(\ILC)bb'} 
\ee
with undetermined $A_1, A_2$ and construct the model correlation matrix 
{$R^{XY,WZ}_{(\mathrm{model})bb'}$}
by generalizing
\eqref{model_correlation}.

We can quantify the level of agreement in the amplitudes by
determining the values $A_1, A_2$ that minimize the residuals between the
correlation matrix from simulations and the model,
\be
\label{A1_A2_fit}
	\sum_{\substack{XY,WZ\\|b - b'| > 2}} \(R^{XY,WZ}_{(\mathrm{model})bb'} - {\hat
	R^{XY,WZ}_{bb'}}\)^2 ,
\ee
and their uncertainties by bootstrap resampling with replacement from our 2400 simulations.  We
consider only multipoles up to $\ell = 3000$ in the minimization. 

Allowing $A_i$ to vary does not substantially decrease the
residuals plotted in Fig.~\ref{fig:fit_residuals}.
The resulting values of $A_i$ with the bootstrapped error bars are shown in
Fig.~\ref{fig:A1_A2} for all five window functions. While the ILC amplitudes are in good
agreement with the theoretical expectation $A_2 = 1$, the SSC amplitudes show a clear
positive bias. Due to super-sample lensing, the acoustic peaks thus shift around their
fiducial positions {slightly} more than predicted by \eqref{sigma_kappa_2},
though this does not have significant bearing on cosmological inferences, see
\S\,\ref{sec:ssc}.  One possible explanation of this discrepancy is complications
arising from the edge effects, not considered in the derivation of \eqref{sigma_kappa_2}.

\begin{figure*}
\center
\includegraphics[width = 0.49 \textwidth]{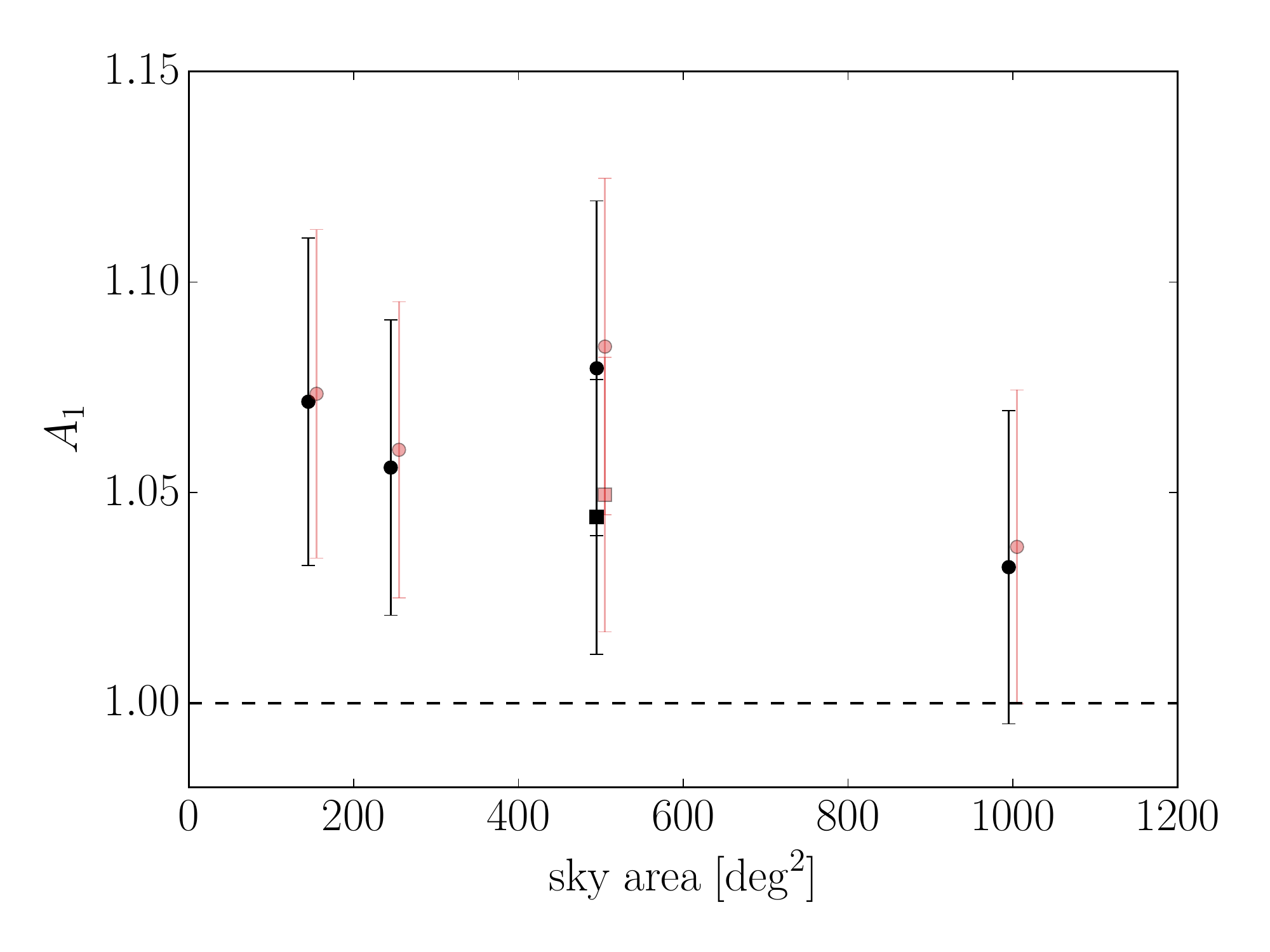}
\includegraphics[width = 0.49 \textwidth]{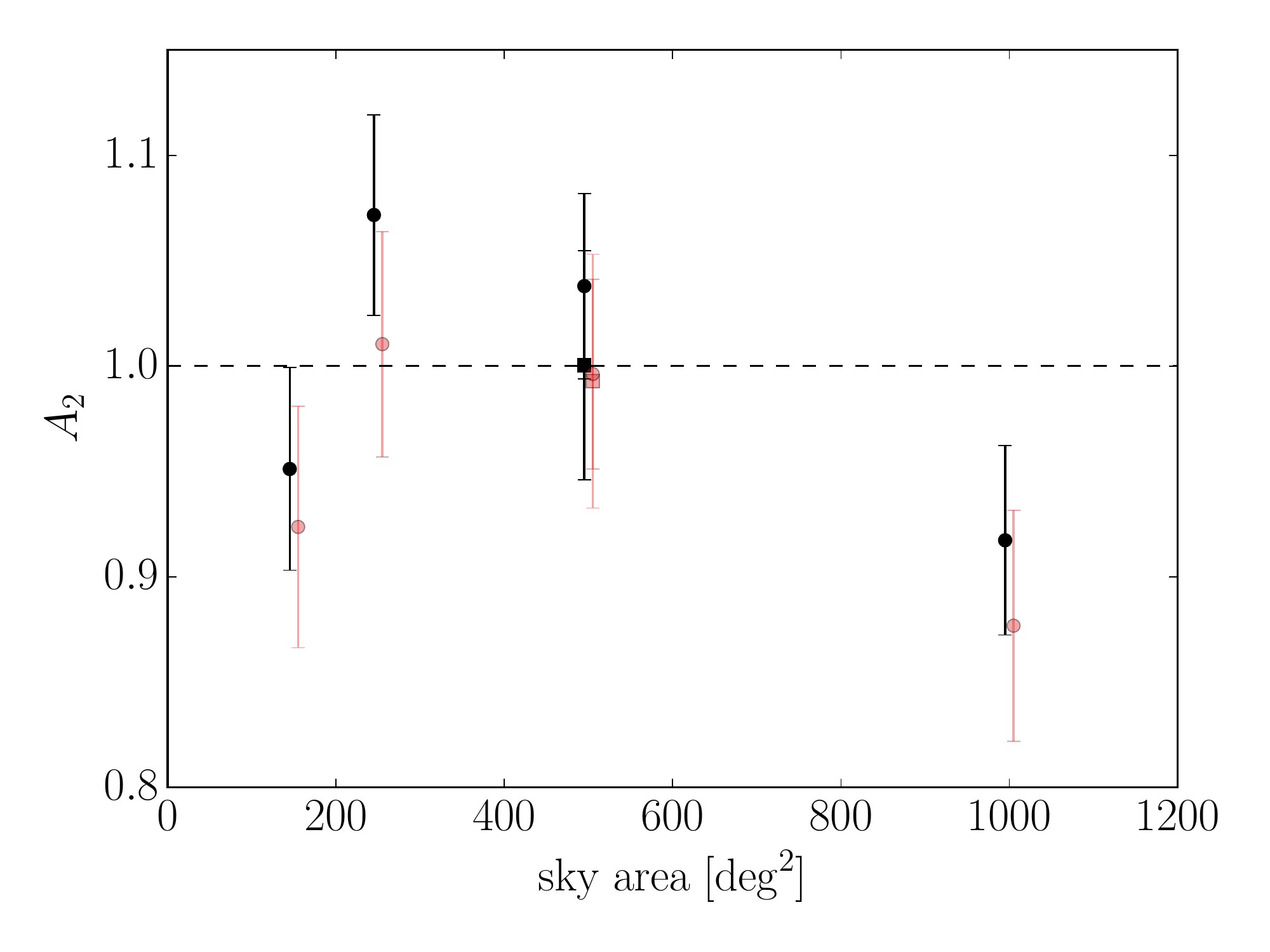}
\caption{Constraints on the amplitude parameters $A_1, A_2$ parameterizing the lensing
induced covariance effects in the full covariance matrix \eqref{model_covariance}, as
determined from the simulations. Black symbols represent values obtained by minimization
of \eqref{A1_A2_fit}, red symbols values obtained using the alternative method from the
Appendix~\ref{sec:app_alternative_method}. Each point represents a different window
function, the circles denote disk-shaped window functions and the squares the SPT-like
rectangular window function. Error bars show 68\% confidence limits obtained by
bootstrapping. Theoretical expectations $A_i = 1$ are marked with the dashed line.} 
\label{fig:A1_A2}
\end{figure*}

\subsection{{Empirical} determination of SSC, ILC {modes}}
\label{sec:app_alternative_method}

In this section we present an alternative method to assess how well the 
covariance matrix for the binned full sky power spectra $\hat{\mathcal{C}}_b^{XY}$
estimated from the simulations agree with the theoretical predictions of the
lensing-induced effects.

The method extends the ideas presented in \cite{HarnoisDeraps:2011rb}, where the 
aim was to parameterize features found in a correlation matrix $R_{ij}$ obtained from
an ensemble of simulations $\hat R_{ij}$ and to decrease the numerical noise due to the finite number of simulations.
The authors assumed that beyond the
diagonal elements, which  are equal to one by definition, the correlation matrix is
relatively smooth. Their analysis proceeds as
follows:
\begin{enumerate}
	\item Initialize the $k=0$ step by setting $R^{k}_{ij}={\hat R_{ij}}$ for $i\ne j$ and
	$R^k_{ij}=0$ for $i=j$. 	
		\item Decompose $R^k_{ij}$  into orthonormal eigenmodes as
		\be
		\label{eigenmode_decomposition}
			R^k_{ij} = \sum_K \lambda^k_K v^k_{K,i} v^k_{K,j}. 
		\ee
		{Examine the eigenvalues for a break in the spectrum
	and identify the $N$ signal dominated modes (see
	below for an example). }
	\item Set $R^{k+1}_{ij}= \hat R_{ij}$ for $i\ne j$ and update its diagonal 
	using  the contribution of the $N$ signal eigenmodes from step 2,
		\be
		\label{ueli_replacement}
			R^{k+1}_{ii} = \sum_{K = 1}^N \lambda^k_K \(v^k_{K,i}\)^2 .
		\ee
	
	\item Repeat steps 2 and 3 with $k\rightarrow k+1$ until the elements on the
	diagonal converge to the required precision\footnote{For the covariance matrices
	investigated in this work we find that relative error on $\lambda_{K, i}^k, v^k_{K,i}$
	for $K = 1,2$ drops by about a factor of three with each additional iteration. We
	performed 15 iterations for each covariance matrix.}.
	\item Approximate the correlation matrix with the eigenvalues and
	eigenvectors of 
	$R^{k_{\rm max}}_{ij}$ as
\be
\label{ueli_final}
	R_{ij} \approx
	\begin{cases}
	 \sum_{K = 1}^N \lambda_K^{k_\mathrm{max}} v_{K,i}^{k_\mathrm{max}} v_{K,j}^{k_\mathrm{max}}\ & i \neq j \\
	 1 & i=j 
	 \end{cases},
\ee

\end{enumerate}

 Because of the
convergence of $\lambda_K^k$ and $v_{K,i}^k$, the diagonal elements \eqref{ueli_replacement}
themselves converge and the estimate of the off-diagonal structure is not biased by the initial omission of the
diagonal.

In our case, a similar procedure can be used with only a small alteration. Unlike in
\cite{HarnoisDeraps:2011rb}, in the correlation matrix \eqref{correlation} there are
features we are not interested in probing not only on the main diagonal $R^{XY,XY}_{bb}$,
but due to Gaussian covariance terms also on the sub-diagonals $R^{XY,WZ}_{bb}$ and due
to the window function effects also on the neighboring bins $R^{XY,WZ}_{bb'}$.  In practice, 
we generalize the procedure  by zeroing out the $b-2 \le b' \le b+2$ elements for all
$XY,WZ$  and replacing them with the iterative construction above.    This is conservative
as we only see evidence for window function effects in the nearest neighboring bin with the fiducial bin
width and windows.  With this procedure, we
isolate the lensing-induced features of the correlation matrix.

In analyzing the simulated covariance matrices ${\hat R^{XY,WZ}_{bb'}}$ we find that with 2400
simulations we can only detect two features -- the SSC and ILC effects. This is illustrated
in Fig.~\ref{fig:ueli_eigenvalues}, where we show distribution of converged eigenvalues
$\lambda_K$ for the 1000 deg${}^2$ disk window when using $N = 2$ in the algorithm above.
The two eigenvalues corresponding to SSC and ILC are clearly separated from the other
eigenvalues.  We infer that the others are too small to be detectable with 2400 simulations. When using other values of $N$,
Fig.~\ref{fig:ueli_eigenvalues} or other results further below do not appreciably change,
we thus quote results for $N=2$ in what follows. The situation is identical for the other
window functions.  

\begin{figure}
\center
\includegraphics[width = 0.49 \textwidth]{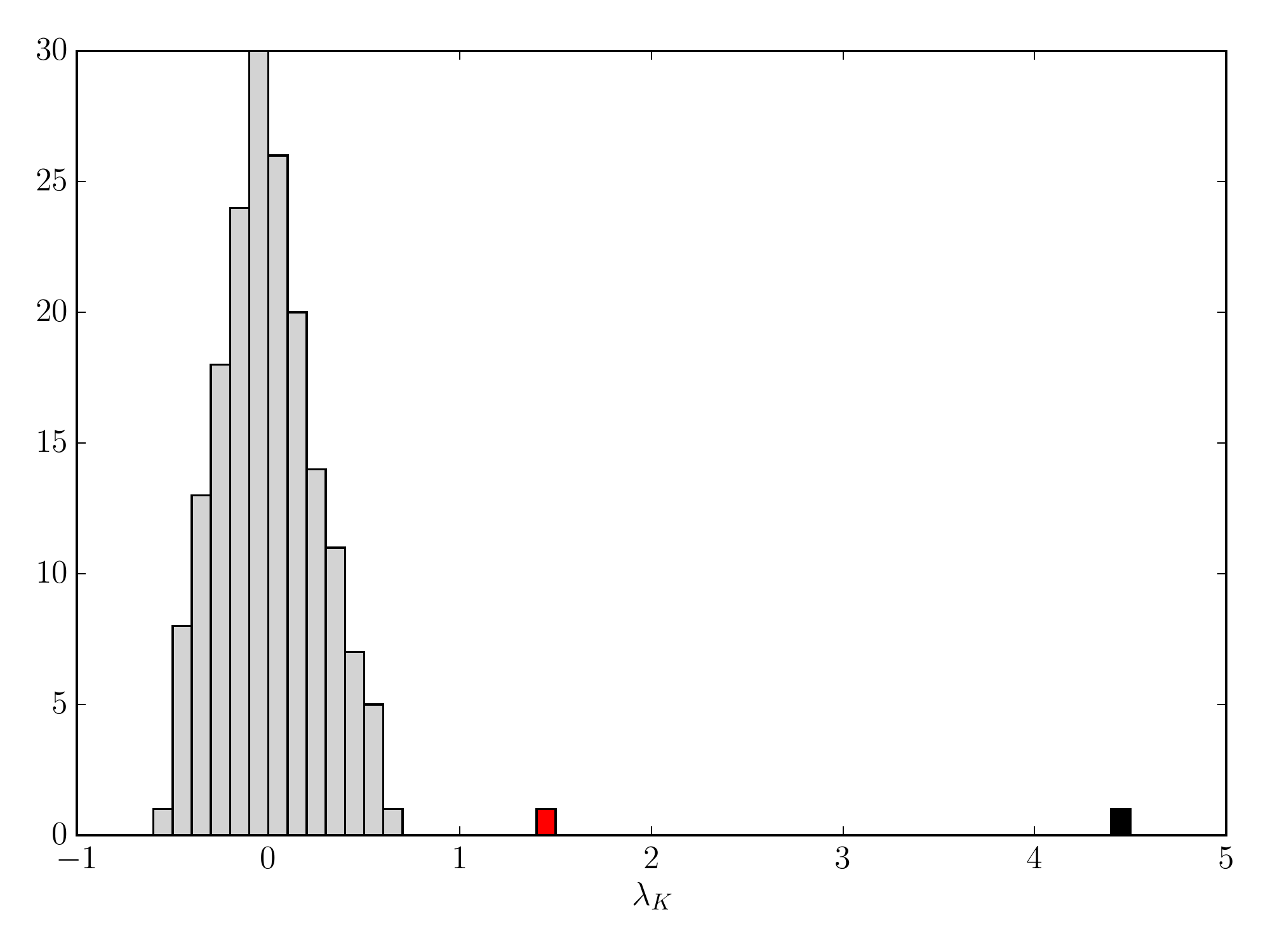}
\caption{Distribution of eigenvalues $\lambda_K$ obtained from the simulated covariance
for the 1000 deg${}^2$ disk window using the method described in
Appendix~\ref{sec:app_alternative_method}. The eigenvalue corresponding to the SSC effect
is colored in black, while the one corresponding to the ILC effect in red. The other
eigenvalues arise from numerical noise and potentially other features of the covariance
matrix that are not distinguishable with 2400 simulations.}
\label{fig:ueli_eigenvalues}
\end{figure}

For each window function, we use the iterative procedure to find vectors $V_1, V_2$ that
approximate the lensing-induced structure of the correlation matrix as
\be
\label{ueli_approx}
	R^{XY,WZ}_{bb'} \approx V^{XY}_{1b}V^{WZ}_{1b'} + V^{XY}_{2b}V^{WZ}_{2b'}
	\ \ \(|b - b'| > 2\).
\ee
These vectors are related to the expansion \eqref{ueli_final} through
\be
	V^{XY}_{Kb} = \sqrt{\lambda^{k_\mathrm{max}}_K} v^{k_\mathrm{max},XY}_{Kb} .
\ee
In Fig.~\ref{fig:fit_residuals}, we {compare the residuals between the simulations and \eqref{ueli_approx} (right) 
with those between the simulations and the theory from the previous section (left).
The residuals for the two
approaches are qualitatively similar with a slightly higher residuals for the latter as would be expected from a
theoretical as opposed to phenomenological model.
}

We can gain further insight about the relationship between the first two eigenmodes
$V^{XY}_{1,2}$ and the theoretical expectations of the SSC and ILC effects by examining 
eigenvectors of the latter.   From \eqref{ssc} we see the contribution of
the SSC term to the correlation matrix can be factored into
\be
	\frac{
	\Cov^{XY,WZ}_{(\SSC)bb'}
	}{
	{\sqrt{\Cov_{bb}^{XY,XY}\Cov_{b'b'}^{WZ,WZ}}} .
	} = U^{XY}_{1b}U^{WZ}_{1b'}
\ee
with
\be
\label{U1}
	U_{1b}^{XY} = 
	\sum_{\ell}
	\frac{\mathcal{U}^{XY}_{b\ell}}{\sqrt{\Cov_{bb}^{XY,XY}}}
	\frac{\partial \ell^2 C^{XY}_\ell}{\partial \ln \ell}
	\frac{\sigma_\kappa}{\ell^2} .
\ee
The ILC covariance including only TT, TE and EE is well captured by a single
eigenvector \cite{Peloton:2016kbw}. Its contribution to the  correlation matrix can be then well approximated by 
\be
	\frac{\Cov^{XY,WZ}_{(\ILC)bb'}}
	{
	{\sqrt{\Cov_{bb}^{XY,XY}\Cov_{b'b'}^{WZ,WZ}}} .
	}\approx U^{XY}_{2b}U^{WZ}_{2b'} 
\ee
for some $U_{2b}^{WZ}$ that can be obtained using eigenvalue decomposition.
Because the leading lensing principal component $\Theta^{(1)}$ captures most of the
lensing effect, $U_2^{XY}$ can be well approximated by
\be
\label{U2_approx}
	U_{2b}^{XY} \approx
	\sum_{\ell}
	\frac{\mathcal{U}^{XY}_{b\ell}}{\sqrt{\Cov_{bb}^{XY,XY}}}
	\frac{\partial  C^{XY}_\ell}{\partial \Theta^{(1)}}
	\sigma_{\Theta^{(1)}}^A ,
\ee
where $\sigma_{\Theta^{(1)}}^A$ is the sample variance of $\Theta^{(1)}$ in the footprint
\eqref{cv_sigma_Th1}. We checked \eqref{U2_approx} explicitly,
but use the numerically obtained value $U_2^{XY}$ in what follows.

If the theoretical predictions for ILC and SSC are correct, we expect
\be
	V^{XY}_{Kb} \approx U^{XY}_{Kb},\ K = 1, 2 .
\ee
In Figure~\ref{fig:evecs_comparison} we compare these two vectors and find that the
agreement is indeed very good, confirming in a different way that our theoretical
understanding of the lensing-induced terms in the simulated covariance matrix is
satisfactory.

\begin{figure}
\center
\includegraphics[width = 0.49 \textwidth]{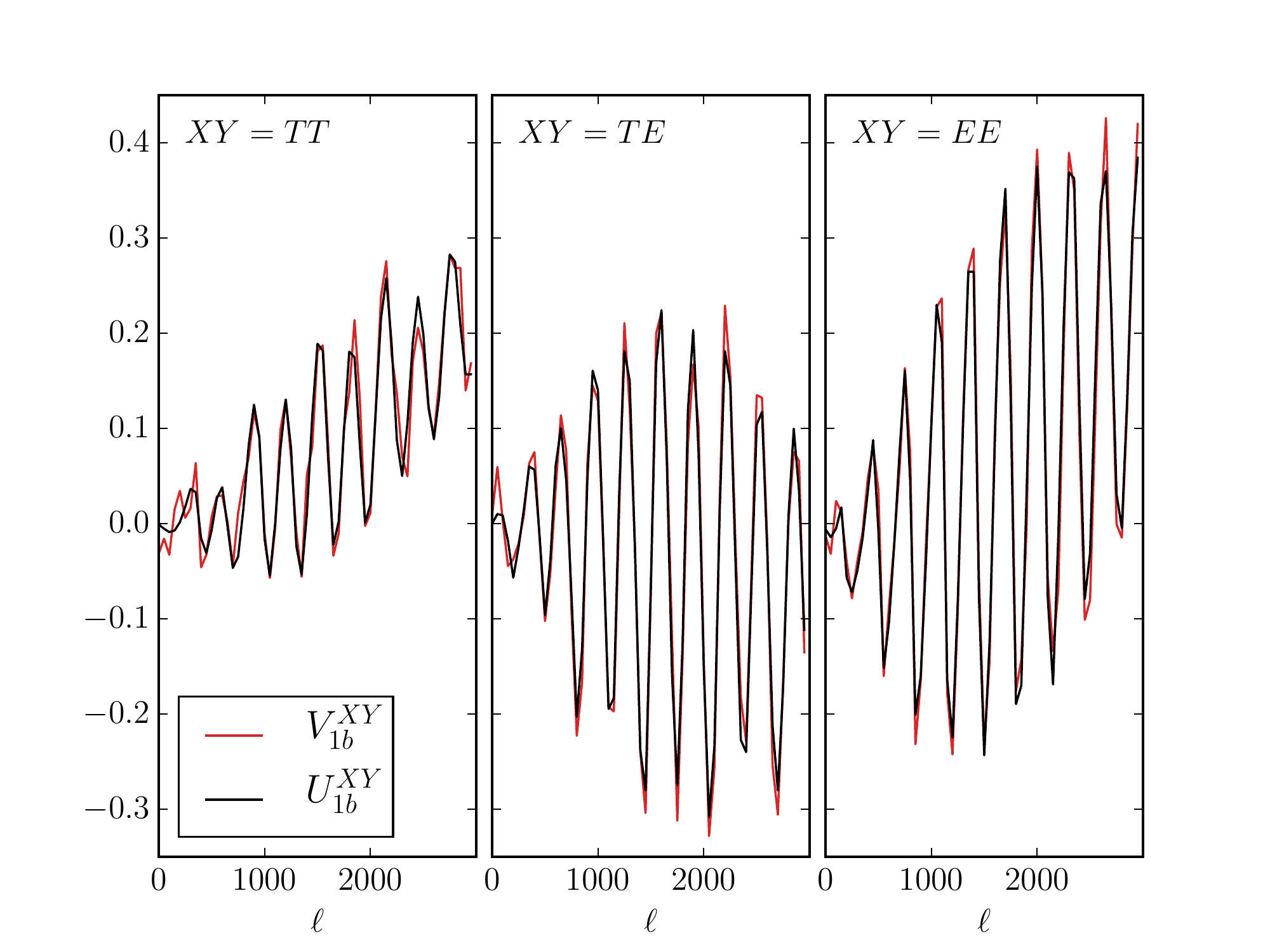}
\includegraphics[width = 0.49 \textwidth]{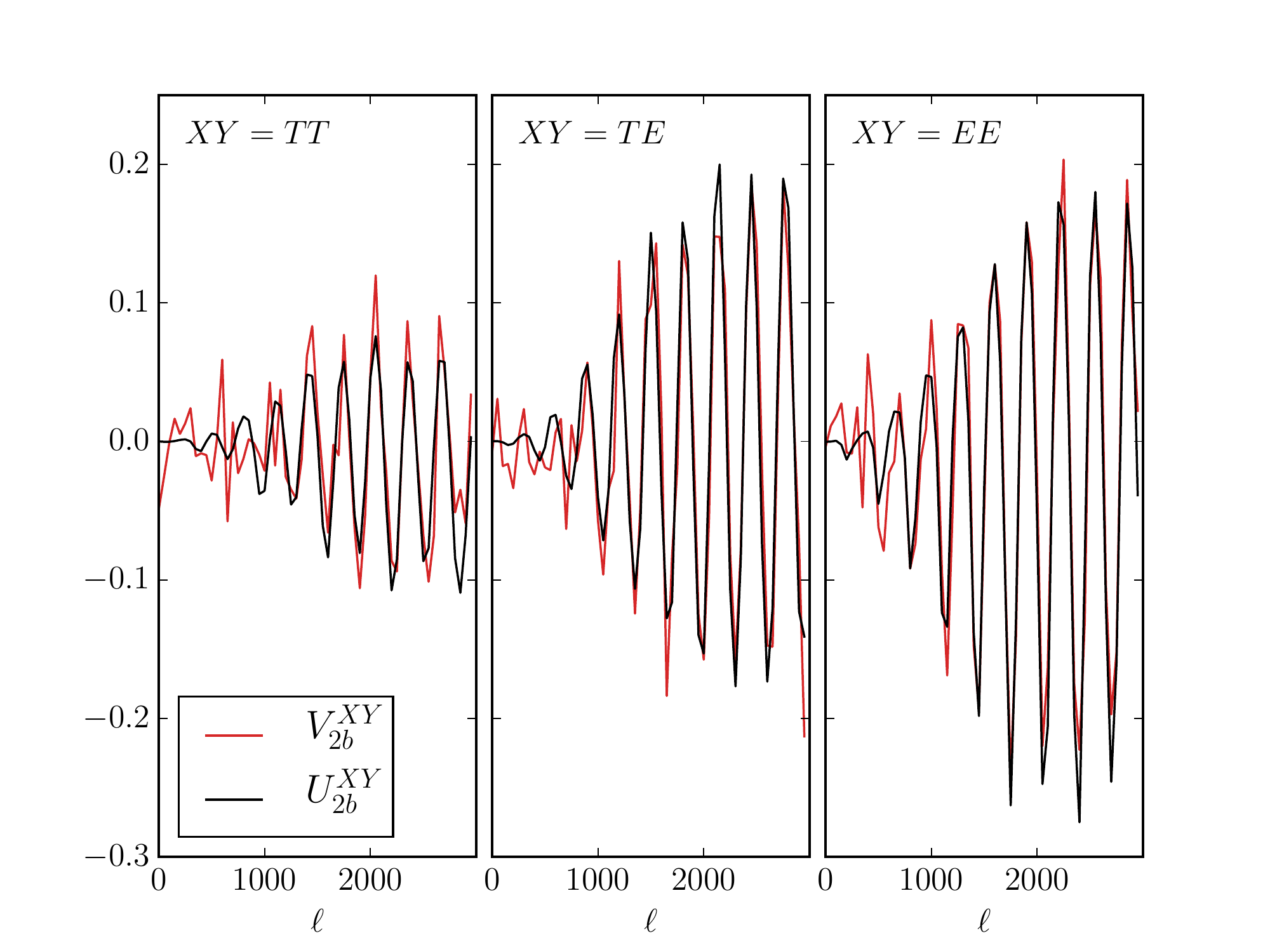}
\caption{The two leading scaled eigenmodes $V^{XY}_{Kb}$ (for $K = 1,2$)
determined from the simulated covariance matrix using the method of
Appendix~\ref{sec:app_alternative_method} (red) against the theoretical expectations
(black). The top plot shows agreement for the SSC effect, while the lower for the ILC
effect. Simulated results were obtained for the 1000 deg${}^2$ disk window function.} 
\label{fig:evecs_comparison} 
\end{figure}

Similarly to before, we can introduce amplitudes $A'_{1,2}$ to quantify agreement between
$V^{XY}_{Kb}$ and $U^{XY}_{Kb}$ and estimate their values by minimizing
\be
	\sum_{XY, b} \(V^{XY}_{1b} - \sqrt{A'_1}U^{XY}_{1b}\)^2
\ee
and similarly for $A'_2$. The square root is used so that in an ideal case, where the two
techniques for comparing simulated and theoretical covariance matrices produce identical
results, we would get $A_K = A_K'$.  We can estimate uncertainties on $A'_{1,2}$ again
using bootstrap. The results are plotted in Fig.~\ref{fig:A1_A2}, where they can be
compared against results of the method introduced in the
Appendix~\ref{sec:app_template_fitting}. The difference between
the two methods is much smaller than the uncertainties for the amplitude of the SSC term
$A_1$, but there seem to be a small bias for $A_2$, with the method presented in
this section obtaining smaller values.

Resolution of this issue could lie in the relative detection significance of these
two effects. From Fig.~\ref{fig:ueli_eigenvalues} it is clear that the SSC
effect is detected with a higher significance and that the eigenvalue corresponding to the
ILC effect is notably closer to the noise-fitting eigenvalues clustered around zero. Our
hypothesis is that $V_2$ obtained from the simulations contains an admixture of the noisy
modes.  This then leads to misalignment of $V_2$ and $U_2$, and a decrease in $A_2'$.
This hypothesis is corroborated by the fact that the agreement between the two methods of
obtaining $A_2$ improves when we drop the temperature data, where the ILC signal is
relatively weaker. Additionally, the difference between the two methods increases when the
analysis is repeated with a smaller number of simulations, in which case we expect larger
admixture of the noisy modes into $V_2$.

Overall, the differences in the amplitudes $A_i$ of ILC and SSC determined from
simulations from the theoretical expectation of unity are small and so in the main text we
simply set $A_i = 1$.

\section{Sharpening of {windowed peaks}}
\label{sec:app_sharpening}

As explained in the main text, we expect cut sky power spectra to exhibit sharper peaks
than the full sky power spectra, because lenses larger than the survey footprint do not
average out to cause peak smoothing, but act as a coherent (de)magnification. Since this
is also the sense in which the SPT data are in tension with Planck, in this Appendix we
demonstrate that it cannot reduce the tension.

To do that, we compare the unbiased estimates of the binned power spectra
$\hat{\mathcal{C}}_b^{XY}$ against the theoretically expected value
$\mathcal{U}^{XY}_{b\ell} C^{XY}_{\ell}$.
Here $C^{XY}_{\ell}$ are the theoretically predicted full sky power spectra
for our fiducial cosmological model. If the amount of lensing in
$\hat{\mathcal{C}}_b^{XY}$ is indeed smaller, it should be possible to detect
nonzero $\Theta^{(1)}$ in the difference
\be
	\Delta C_b^{XY} = \hat{\mathcal{C}}_b^{XY} - \sum_\ell
	\mathcal{U}^{XY}_{b\ell} C^{XY}_{\ell} .
\ee

We model this difference as
\be
	\Delta C_{(\mathrm{model})b}^{XY} = 
	\sum_{\ell}\mathcal{U}^{XY}_{b\ell} \(\Delta \Theta^{(1)}
	\frac{\partial C^{XY}_{\ell}}{\partial \Theta^{(1)}}
	+ \Delta \theta_* \frac{\partial C^{XY}_{\ell}}{\partial \theta_*}\)
	,
\ee
{to account for} the two effects lensing has on power spectra in a cut sky experiment.

From the simulated $\Delta C_b^{XY}$ we then constrain $\Delta \Theta^{(1)}, \Delta
\theta_*$ for each simulation and each window function by minimizing
\be
	\sum_{\substack{XY,WZ\\bb'}} 
	\Delta_b^{XY}
	\(\Cov^{XY,WZ}_{bb'}\)^{-1}
	\Delta_{b'}^{WZ} ,
\ee
where the residuals are
\be
	\Delta_b^{XY} = \Delta C_b^{XY} - \Delta C_{(\mathrm{model})b}^{XY} ;
\ee
this minimization can be done algebraically. For each window we can then read off mean
values of $\Delta \Theta^{(1)}$ and $\Delta \theta_*$ and their variance from the obtained
distribution.

\begin{figure}
\center
\includegraphics[width = 0.49 \textwidth]{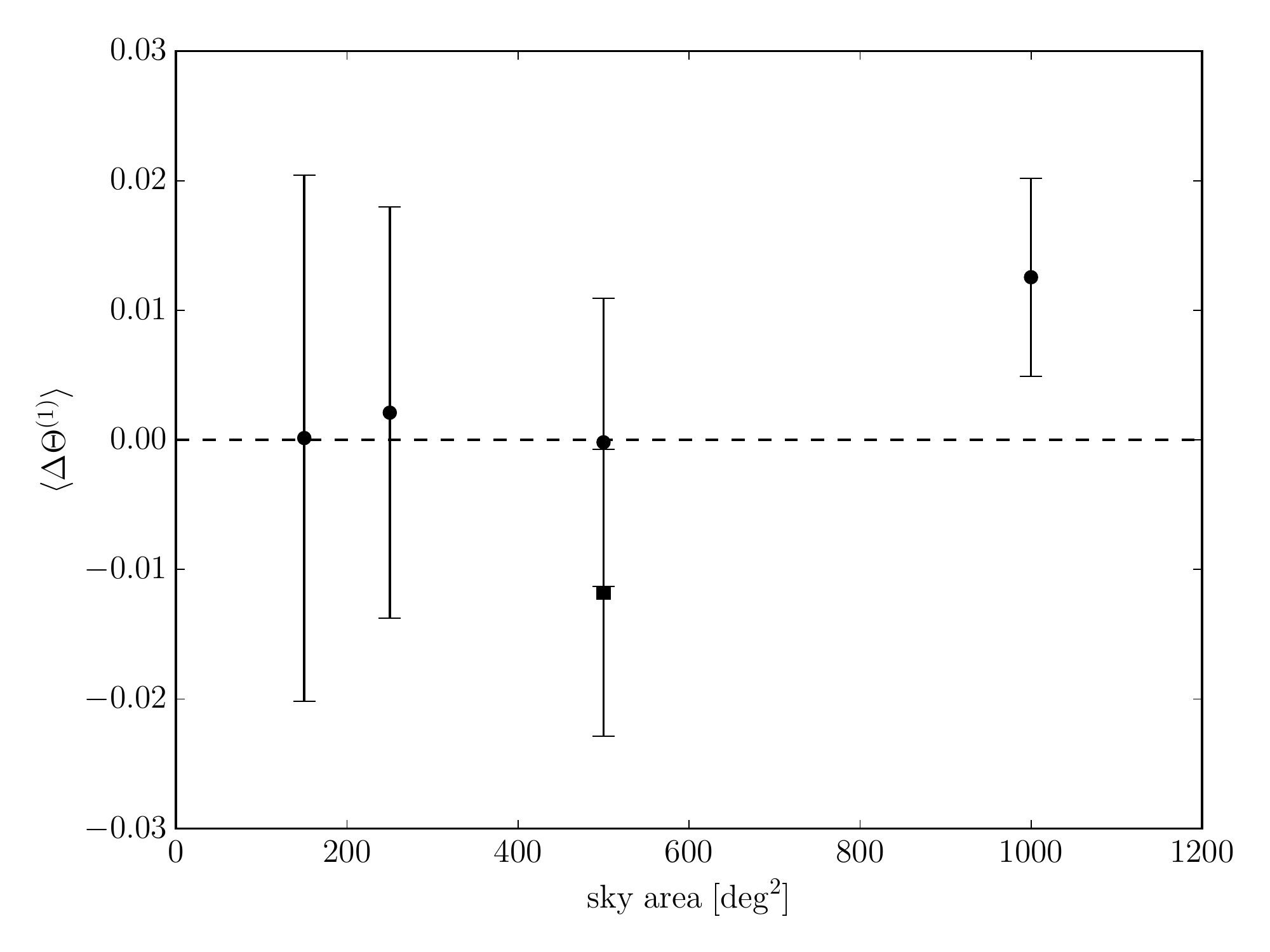}
\caption{Constraints on extra sharpening of the peaks in cut-sky simulations, studied in
Appendix~\ref{sec:app_sharpening}. Each point represents a different window
function, the circles denote disk-shaped window functions and the square the SPT-like
rectangular window function. For each window, the central value represents a mean of
the 2400 values of $\Delta \Theta^{(1)}$ obtained from our simulations. The error bars
represent error on this mean, standard deviation of the $\Delta \Theta^{(1)}$ distribution
divided by $\sqrt{2400}$. 
}
\label{fig:extra_sharpening}
\end{figure}

As expected, mean shift of the peaks $\langle \Delta \theta_*\rangle$ is consistent with zero.
So is the mean shift in $\langle \Delta \Theta^{(1)}\rangle $ that quantifies the extra sharpening of the
peaks; in Fig.~\ref{fig:extra_sharpening} we show constraints on $\langle \Delta \Theta^{(1)}\rangle$ for
all the window functions considered in this work. Even though on theoretical grounds the
sharpening of the peaks is expected to be present, our simulations limit the magnitude of
this effect to be a small fraction of the $\Theta^{(1)}$ standard deviation due to lens
sample variance {\eqref{cv_sigma_Th1}}, i.e. $|\langle \Delta
\Theta^{(1)}\rangle|\ll \sigma_{\Theta^{(1)}}^A$.

\section{Tension significance for non-Gaussian posteriors}
\label{sec:app_significance}

Determining the level of agreement between two measurements of a variable $x$ is an often
encountered problem. The approximation usually employed is to assume independence of these
measurements and approximate the corresponding posterior probability densities $P_1(x),
P_2(x)$ as two Gaussians with means $\mu_i$ and variances $\sigma_i^2$. The tension in the
units of the total variance $\sigma$ is then calculated using the difference of the means
formula
\be
\label{difference_of_the_means}
	T = \frac{|\mu_1 - \mu_2|}{\sqrt{\sigma_1^2 + \sigma_2^2}}.
\ee

In the case where the two posteriors are not Gaussian, it may be possible to apply a
nonlinear transformation of the variable $x$, after which the posteriors are better
approximated by Gaussians. This was the path taken for example in \cite{Motloch:2018pjy}.
Finding such a transformation can be time consuming and may not be always possible. For
that reason we introduce here a tension statistic that can be directly used in the general
case.

Let us assume the two measurements are in principle correlated and described by the 
posterior probability density $P(x_1, x_2)$, where $x_1, x_2$ label results of the two
measurements. For such $P$ we evaluate
\ba
	\mathcal{P}(2 > 1) &=& \int_0^\infty d\Delta \int P(x, x + \Delta) dx \nonumber\\
	\mathcal{P}(1 > 2) &=& \int_{-\infty}^0 d\Delta \int P(x, x + \Delta) dx  \nonumber\\
	&=& 1 - \mathcal{P}(2 > 1) 
\ea
and calculate $T$ from
\be
\label{tension_formula}
	\min \big[ \mathcal{P}(2 > 1), \mathcal{P}(1 > 2) \big] = 
	\int_T^\infty 
	\frac{{d}x}{\sqrt{2\pi}}
	e^{-{x^2}/{2}} .
\ee
The tension significance is then marked as $T \sigma$.

For the special case where $P$ factorizes into two independent Gaussians this definition
is equivalent to the standard formula \eqref{difference_of_the_means}. Additionally, this 
tension significance is invariant with respect to reparameterizations of the measured
variable $x$.

\bibliography{sptlens}

\end{document}